\documentclass[twocolumn,showpacs,aps,prd,superscriptaddress]{revtex4}
\usepackage{dcolumn}
\usepackage{amsmath}

\usepackage{graphicx}

\newcommand{\BABARPubYear}    {05}
\newcommand{\BABARPubNumber}  {45}
\newcommand{\SLACPubNumber} {11499}
\RequirePackage{xspace}
\usepackage{relsize}
\def\babar{\mbox{\slshape B\kern-0.1em{\smaller A}\kern-0.1em
    B\kern-0.1em{\smaller A\kern-0.2em R}}}

\def\Bbar    {\kern 0.18em\overline{\kern -0.18em B}{}\xspace}
\def\BB      {\ensuremath{B\Bbar}\xspace}
\def\Y#1S{\ensuremath{\Upsilon{(#1S)}}\xspace}% no space before {...}!
\def\FourS {\Y4S}
\def\jpsi     {\ensuremath{{J\mskip -3mu/\mskip -2mu\psi\mskip 2mu}}\xspace}
\def\BR         {{\ensuremath{\cal B}\xspace}}
\newcommand{\gev}{\ensuremath{\mathrm{\,Ge\kern -0.1em V}}\xspace}
\newcommand{\mev}{\ensuremath{\mathrm{\,Me\kern -0.1em V}}\xspace}
\newcommand{\gevc}{\ensuremath{{\mathrm{\,Ge\kern -0.1em V\!/}c}}\xspace}
\newcommand{\mevc}{\ensuremath{{\mathrm{\,Me\kern -0.1em V\!/}c}}\xspace}
\newcommand{\gevcc}{\ensuremath{{\mathrm{\,Ge\kern -0.1em V\!/}c^2}}\xspace}
\def\ra                 {\ensuremath{\rightarrow}\xspace}
\def\pep2{PEP-II}

\begin{document}
{\pagestyle{empty}

\begin{flushleft}
\babar-PUB-\BABARPubYear/\BABARPubNumber, \,\, \\
SLAC-PUB-\SLACPubNumber \\

\end{flushleft}

% Title of the paper
\title{
\Large \bf Measurement of the Inclusive Electron Spectrum in Charmless
Semileptonic {\boldmath $B$} Decays Near the Kinematic Endpoint and Determination of 
{\boldmath $|V_{ub}|$}
}
\bigskip

\bigskip \bigskip

% Abstract
\begin{abstract}

We present a measurement of the inclusive electron spectrum in 
$B\to X_u e \nu$ decays near 
the kinematic limit for $B\to X_c e \nu$ transitions, using a  
sample of 88 million \BB\ pairs 
recorded by the \babar\ detector 
at the \FourS\ resonance. Partial branching fraction measurements 
are performed in five overlapping intervals of the electron momentum;
for the interval of $2.0\,$--$\,2.6 \gevc$ 
we obtain $\Delta {\cal B}(B\ra X_u e \nu) = 
(0.572 \pm 0.041_{\mathit{stat}} \pm 0.065_{\mathit{syst}})\times 10^{-3}$. 
Combining this result with shape function parameters extracted from \babar\ 
measurements of moments of the inclusive photon spectrum in 
$B\ra X_s \gamma$ decays and moments of the hadron mass and 
lepton energy spectra in $B\ra X_c \ell \nu$ decays we determine $|V_{ub}|= 
(4.44 \pm 0.25_{\mathit{exp}} \,^{+0.42}_{-0.38\,{\mathit{SF}}} \pm 0.22_{\mathit{theory}}) \times 10^{-3}$. 
Here the first error represents  the combined statistical and systematic 
experimental uncertainties of the partial branching fraction measurement,
the second error refers to the uncertainty of the determination of the shape
function parameters, and the third error is due to theoretical uncertainties 
in the QCD calculations.
\vfill

\end{abstract}

\author{B.~Aubert}
\author{R.~Barate}
\author{D.~Boutigny}
\author{F.~Couderc}
\author{Y.~Karyotakis}
\author{J.~P.~Lees}
\author{V.~Poireau}
\author{V.~Tisserand}
\author{A.~Zghiche}
\affiliation{Laboratoire de Physique des Particules, F-74941 Annecy-le-Vieux, France }
\author{E.~Grauges}
\affiliation{IFAE, Universitat Autonoma de Barcelona, E-08193 Bellaterra, Barcelona, Spain }
% ...
\collaboration{The \babar\ Collaboration}
\noaffiliation

\pacs{13.20.He,                 % semileptonic bottom meson decays
      12.15.Hh,                 % CKM elements determination
      12.38.Qk,                 % Experimental tests of QCD calculations
      14.40.Nd}                 % properties of bottom mesons

\maketitle

}

% The body of the paper starts here
\section{Introduction}
\label{sec:introduction}

The Cabibbo-Kobayashi-Maskawa (CKM) matrix element $V_{ub}$,
the coupling of the $b$ quark to the $u$ quark,
is a fundamental parameter of the Standard Model.
It is one of the smallest and least known elements of the CKM matrix.
With the increasingly precise measurements of decay-time-dependent $CP$
asymmetries in $B$-meson decays, in particular the angle
$\beta$~\cite{beta,phi1},
improved measurements of the magnitude of $V_{ub}$ will allow
for stringent experimental tests of the Standard Model
mechanism for $CP$ violation~\cite{sm}. This is best illustrated
in terms of the unitarity triangle, the graphical representation of
the unitarity condition for the CKM matrix, for which the length of
the side that is opposite to the angle $\beta$ is proportional to
$|V_{ub}|$.

The extraction of $|V_{ub}|$ is a challenge, both theoretically and
experimentally.
Experimentally, the principal challenge is to separate the signal
$B \rightarrow X_u e\nu$ decays from the 50 times larger
$B\rightarrow X_c e\nu$ background.
This can be achieved by selecting regions of phase space in which
this background is highly suppressed.
In the rest frame of the $B$ meson, the
kinematic endpoint of the electron spectrum is $\sim2.3 \gevc$
for the dominant $B \rightarrow X_c e \nu$ decays and
$\sim 2.6 \gevc$ for $B \rightarrow X_u e \nu$ decays.
Thus the spectrum above 2.3~\gevc is dominated by electrons
from $B \rightarrow X_u e \nu$ transitions. This allows for
a relatively precise measurement, largely free from \BB\ background,
in a 300 \mevc interval that covers
approximately 10\% of the total electron spectrum for charmless
semileptonic $B$ decays.
In the $\Upsilon(4S)$ rest frame, the finite momenta of the $B$ mesons cause
additional spread of the electron momenta of $\sim 200$\mevc, extending the
endpoints to higher momenta.

The weak decay rate for $b\ra u e \nu$ can be calculated
at the parton level. It is proportional to $|V_{ub}|^2$ and $m_b^5$,
where $m_b$ refers to the $b$-quark mass. To relate the semileptonic decay
rate of the $B$ meson to $|V_{ub}|$, the parton-level calculations have to
be corrected for perturbative
and non-perturbative QCD effects. These corrections can be calculated
using various techniques: heavy quark expansions (HQE)~\cite{ope}
and QCD factorization~\cite{neubert94}.
Both approaches separate perturbative from non-perturbative expressions
and sort terms in powers of $1/m_b$.
HQE is appropriate for the calculations of total inclusive $B$ decay rates and
for partial $B$ decay rates integrated over sufficiently large regions of
phase space where the mass and momentum of the final state hadron are
large compared to $\Lambda_{\mathit{QCD}}$.
QCD factorization is better suited for calculations of partial rates and
spectra near the kinematic boundaries where the hadronic mass is small.
In this region the spectra
are affected by the distribution of the $b$-quark momentum inside the
$B$ meson~\cite{neubert94}, which can be described by a structure or shape
function (SF), in addition to weak annihilation and other non-perturbative
effects.
Extrapolation from the limited momentum range near the endpoint
to the full spectrum is a difficult task, because the SF cannot be calculated.
To leading order, the SF should be universal for all
$b \ra q $ transitions (here $q$ represents a light
quark)~\cite{kolya94,neubert94a}.
Several functional forms for the SF, which generally depend on two 
parameters related to 
the mass and kinetic energy of the $b$-quark, $\bar{\Lambda}$ 
or $m_b$, and $\lambda_1$ or $\mu_{\pi}^2$, have been proposed.
The values and precise
definitions of these parameters depend on the specific ansatz for the SF,
the mass renormalization
scheme, and the renormalization scale chosen.

In this paper, we present a measurement of the inclusive electron momentum
spectrum in charmless semileptonic $B$ decays, averaged over charged and
neutral $B$ mesons, near the kinematic endpoint.
We report measurements of the partial branching fractions in five
overlapping momentum intervals. The upper limit is fixed at 2.6 \gevc, while
the lower limit varies from 2.0 \gevc to 2.4 \gevc.
By extending the interval for the signal extraction down to 2.0 \gevc,
we capture about 25\% of the total signal electron spectrum,
but also much larger $B\ra X_c e \nu$ backgrounds.
Inclusive measurements of $|V_{ub}|$ have been performed by several
experiments operating at the $\FourS$ resonance, namely ARGUS~\cite{argus},
CLEO~\cite{cleo,cleo2}, \babar~\cite{ichep04},
and Belle~\cite{belle-vub}, and experiments operating at the
$Z^0$ resonance, namely L3~\cite{l3-vub}, ALEPH~\cite{aleph-vub},
DELPHI~\cite{delphi-vub}, and OPAL~\cite{opal-vub}.
This analysis is based on a method similar to the one used in previous measurements
of the lepton spectrum near the kinematic endpoint~\cite{argus,cleo}.
The results presented here supersede those of the preliminary analysis
reported by the \babar\ Collaboration~\cite{ichep04}.

The extraction of $|V_{ub}|$ relies on two different
theoretical calculations of the differential decay rates for $B\ra X_u e \nu$ and
$B\ra X_s \gamma$:
the original work by DeFazio and Neubert (DN)~\cite{dFN}, and
Kagan and Neubert~\cite{kagan_neubert},
and the more comprehensive recent calculations by Bosch, Lange,
Neubert, and Paz (BLNP)~\cite{blnp1,blnp2,blnp3,blnp4,neubert_loops,neubert_extract}.

The DN calculations allow for the extrapolation of the observed
partial $B\to X_u e \nu$ decay rate above a certain electron
momentum to the total inclusive $B\to X_u e \nu$ decay rate using
the measured SF parameters and a subsequent translation of the
total decay rate to $|V_{ub}|$.
The theoretical uncertainties on the rate predictions are estimated to be of order 10--20\%.

The BLNP authors have presented a systematic treatment of the SF effects,
incorporated all known corrections to the differential decay rates,
and provided an interpolation between the HQE and the SF regions.
They have also performed a detailed analysis of the theoretical uncertainties.
The calculations directly relate the partial decay rate to $|V_{ub}|$.
While the calculations by BLNP are to supersede the earlier work by DN, we use
both approaches to allow for a direct comparison of the two calculations,
and also a comparison with previous measurements based on the DN calculations.
We adopt the SF parameters extracted by the \babar\ Collaboration:
for the DN method we rely
on the photon spectrum in $B\ra X_s \gamma$ decays~\cite{babar_photons};
for the more recent BLNP method, we also use SF parameters derived from
the photon spectrum, its moments, the hadron-mass and lepton-energy
moments in inclusive $B\ra X_c \ell \nu$ decays~\cite{babarvcb}, and the
combination of all moments measured by the \babar\ Collaboration~\cite{babarcf}.

\section{Data Sample, Detector, and Simulation}
\label{sec:Detector}

The data used in this analysis were recorded with the
\babar\ detector at the \pep2 energy-asymmetric $e^+e^-$ collider.
The data sample of 88 million \BB\ events,
corresponding to an integrated luminosity
of 80.4~$\mathrm{fb}^{-1}$, was collected at the \FourS\ resonance.
An additional sample of 9.5~$\mathrm{fb}^{-1}$ was recorded
at a center-of-mass (c.m.) energy 40 \mev below the \FourS\ resonance, {\it i.e.} just
below the threshold for \BB\ production.
This off-resonance data sample is used to subtract the non-\BB\ contributions
from the data collected on the \FourS\ resonance.
The relative normalization of the two data samples
has been derived from luminosity
measurements, which are based on the number of detected $\mu^+\mu^-$
pairs and the QED cross section for
$e^+e^-\to \mu^+\mu^-$
production, adjusted for the small difference in center-of-mass energy.

The \babar\ detector has been described in detail elsewhere \cite{detector}.
The most important components for this study are the charged-particle
tracking system, consisting of a five-layer
silicon detector and a 40-layer drift chamber, and the electromagnetic
calorimeter assembled from
6580 CsI(Tl) crystals. These detector components operate in a
$1.5$-$\mathrm{T}$ solenoidal magnetic field. Electron candidates are
selected on the
basis of the ratio of the energy detected in the calorimeter
to the track momentum, the calorimeter shower shape,
the energy loss in the drift chamber, and the angle of the photons
reconstructed in a ring-imaging Cherenkov detector.

The electron identification efficiency and the probabilities to misidentify
a pion, kaon, or proton as an electron have been measured~\cite{thorsten}
as a function of the laboratory momentum and angles with clean samples of
tracks selected from data.
Within the acceptance of the calorimeter, defined by the polar angle
in the laboratory frame, $-0.72 < \cos \theta_{lab} < 0.92$, the average
electron identification efficiency is $92\%$. The average hadron
misidentification rate
is about 0.1\%.

We use Monte Carlo (MC) techniques to simulate the production
and decay of $B$ mesons, and the detector response~\cite{geant4}, to estimate signal and background efficiencies,
and to extract the observed signal and background distributions. The simulated
sample of generic \BB\ events exceeds the \BB\ data sample by
about a factor of three.

Information from studies of selected control data samples on
efficiencies and resolutions is used to improve the accuracy of the simulation.
Comparisons of data with the MC simulations have revealed small differences in the tracking efficiencies,
which have been corrected for. No significant impact of non-Gaussian
resolution tails has been found for high momentum tracks in the endpoint region.
The MC simulations include radiative effects such as bremsstrahlung
in the detector material and QED initial and final state
radiation~\cite{photos}. Adjustments for small variations of the beam
energy over time have also been included.

In the MC simulations the branching fractions for hadronic
$B$ and $D$ decays are based on values reported in the
Review of Particle Physics~\cite{pdg2002}.
The simulation of charmless semileptonic decays, $B \ra X_u e \nu$, is
based on a heavy quark expansion to 
${\cal O}(\alpha_s)$~\cite{dFN}. This calculation produces a
continuous spectrum of hadronic states. The
hadronization of $X_u$ with masses above $2 m_{\pi}$
is performed by JETSET~\cite{jetset}. The motion
of the $b$ quark inside the $B$ meson is implemented with the
SF parameterization given in~\cite{dFN}.
Three-body decays to low-mass hadrons,
$(X_u = \pi, \rho, \omega,\eta,\eta ')$, are simulated separately
using the ISGW2 model~\cite{isgw2} and mixed with
decays to non-resonant and higher mass resonant states $X^*_u$, so that the
cumulative distributions of the hadron mass, the momentum transfer squared,
and the electron momentum reproduce the HQE calculation as closely as possible.
The generated electron spectrum is reweighted
to accommodate variations due to specific choices of the SF parameters.

\begin{figure}
\begin{center}
\includegraphics[height=6.5cm]{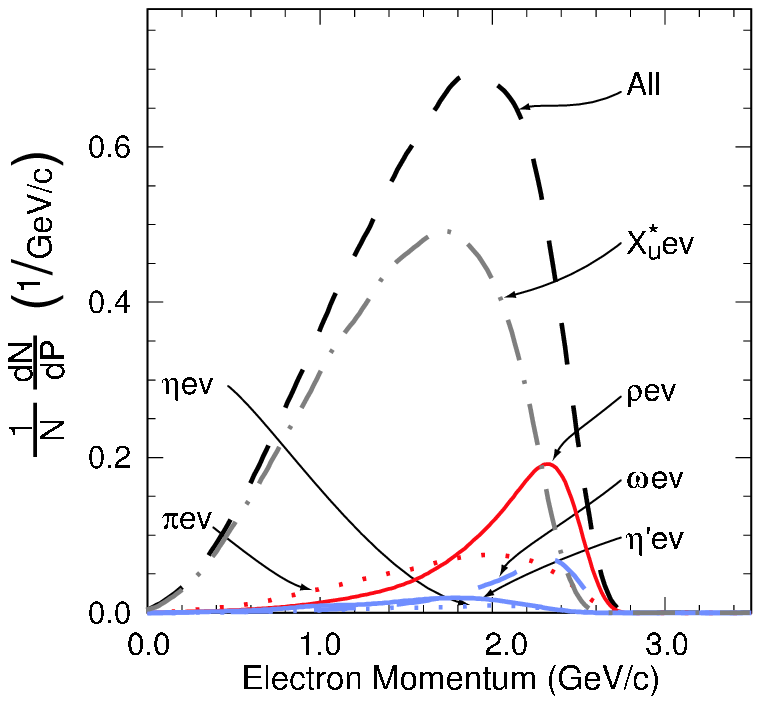}
\caption{
MC-generated electron momentum spectra for various charmless semileptonic
$B$ decays: $B \rightarrow \pi e \nu$, $B \rightarrow \rho e \nu$,
$B \rightarrow \omega e \nu$, $B \rightarrow \eta e \nu$, $B \rightarrow \eta^{\prime} e \nu$,
the sum of $B$-meson decay modes to non-resonant and higher-mass resonance states ($X^*_ue\nu$),
and the sum of all decay modes (All). The spectra are normalized to a total rate of 1.0.
}
\label{f:bu_pe}
\end{center}
\end{figure}

The MC-generated electron-momentum distributions for
$B \rightarrow X_u e \nu$ decays are shown in Fig.~\ref{f:bu_pe},
for individual decay modes and for their sum. Here and throughout the paper,
the electron momentum and all other kinematic variables are measured in the
\FourS\ rest frame, unless stated otherwise.
Above 2\gevc, the principal signal contributions are from decays
involving the light mesons $\pi, \rho$, and $\omega$, and also some
higher mass resonant and non-resonant states $X^*_u$.

For the simulation of the dominant $B \ra X_c e \nu$ decays, we have chosen
a variety of models.
For $B\to D e \nu$ and $B \ra D^* e \nu$ decays we use
parameterizations~\cite{hqet,CLN,GL} of the form factors, based on heavy
quark effective theory (HQET).
Decays to pseudoscalar mesons are described by a single form factor
$F_D(w)/F_D(1) = 1 - \rho_D^2 (w-1)$, where the variable $w$ is the scalar product of the
$B$ and $D$ meson four-vector velocities and is equal to
the relativistic boost of the $D$ meson in the $B$ meson rest frame.
The linear slope $\rho_D^2$ has been measured by the CLEO \cite{bd_form} and
Belle \cite{bd_form2} Collaborations. We use the average value,
$\rho_D^2 = 0.72 \pm 0.12$.
The differential decay rate for $B\to D^*e \nu$
can be described by three amplitudes, which depend on three
parameters: $\rho^2$, $R_1$, and $R_2$. We adopt values recently measured by \babar~\cite{babarff}:
$\rho^2=0.769 \pm 0.043 \pm 0.032$,
$R_1=1.328 \pm 0.060 \pm 0.025$,
and $R_2=0.920 \pm 0.048 \pm 0.013$.
Here the parameter $\rho^2$ is the slope assuming a linear dependence of the form
factor on the variable $w$. The quoted errors reflect the statistical and
systematic uncertainties.

We use the ISGW2~\cite{isgw2} model for various decays to higher-mass
$D^{**}$ resonances. We have adopted a prescription
by Goity and Roberts~\cite{gr} for the non-resonant $B \ra D^{(*)} \pi e \nu$
decays.

\begin{figure}[!htb]
\begin{center}
\includegraphics[height=6.5cm]{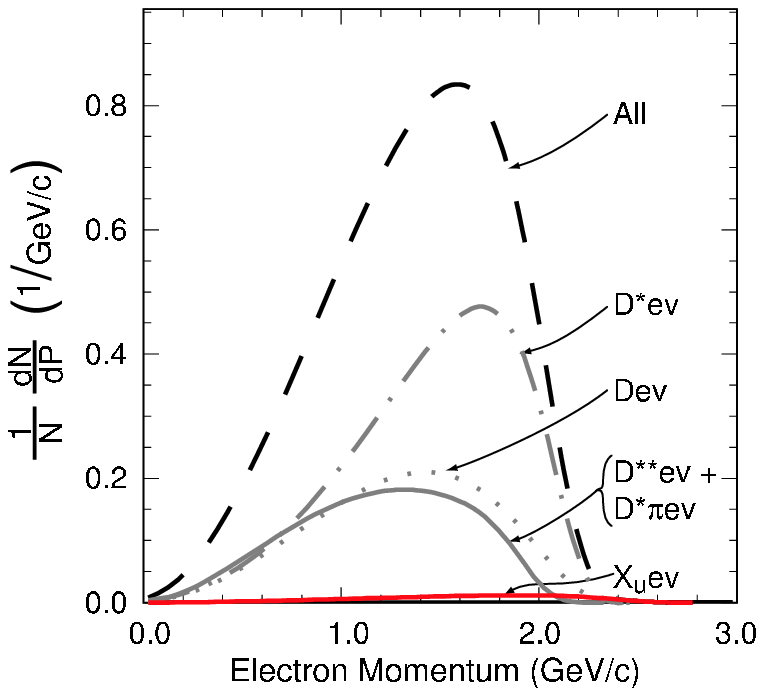}
\caption{
MC-generated electron momentum spectra for
various $B\ra X_c e \nu$ decay modes: $B \rightarrow D e\nu$,
$B \rightarrow D^* e \nu$, $B \rightarrow D^{**}e \nu$,
$B \rightarrow D^{(*)}\pi e \nu$, and $B \rightarrow X_u e \nu$,
and the sum of all decay modes (All). The signal
$B \ra X_u e \nu$ spectrum is shown for comparison.
The spectra are normalized to a total rate of 1.0.
}
\label{f:sp1}
\end{center}
\end{figure}

The shapes of the MC-generated electron spectra for individual
$B \ra X_c e \nu$ decays are shown in Fig.~\ref{f:sp1}.
Above 2 \gevc the principal background contributions are from decays
involving the lower-mass charm mesons, $D^*$ and $D$. Higher-mass and non-resonant
charm states are expected to contribute at lower electron momenta.
The relative contributions of the individual
$B \ra X_c e \nu$ decay modes are adjusted to match the data by a fit to the observed
spectrum (see below).

\section{Analysis}
\label{sec:analysis}

\subsection{Event Selection}
\label{sec:Selection}

We select events with a semileptonic $B$ decay by requiring
an electron with momentum $p_e > 1.1 \gevc$.
To reject electrons from the
decay $\jpsi \ra e^+ e^-$, we combine the electron candidate with any
second electron of opposite charge and reject the combination, if the
invariant mass of the pair falls in the interval
$3.00 < m_{ee} < 3.15 \gevcc$.

To suppress background from non-\BB\ events, primarily
low-multiplicity
QED (including $\tau^+ \tau^-$ pairs) and $e^+e^- \ra q \bar{q}$ processes
(here $q$ represents any of the $u, d, s$ or $c$ quarks), we veto events
with fewer than four charged tracks. We
also require that the ratio of the second to the zeroth Fox-Wolfram
moment~\cite{foxw}, ${\cal R}_2$, not exceed $0.5$. ${\cal R}_2$ is calculated
including all detected charged particles and photons.
For events with an electron in the momentum interval
of 2.0 to 2.6~\gevc, these two criteria reduce the non-\BB\ background
by a factor of about 6, while the loss of signal events is less than 20\%.

In semileptonic $B$ decays, the neutrino carries sizable energy.
In events in which the only undetected particle is
this neutrino, the neutrino four-momentum can be inferred from the missing
momentum, $p_{miss}=(E_{miss},\vec{p}_{miss})$, the difference between the
four-momentum of the two colliding-beam particles, and the sum of the
four-momenta of all detected particles, charged and neutral.
To improve the reconstruction of the missing momentum, we impose a
number of requirements on the charged and neutral particles.
Charged tracks are required to have a minimum transverse
momentum of 0.2 \gevc and a maximum momentum of 10~\gevc in the
laboratory frame. Charged tracks are also restricted in polar angle to
$-0.82 < \cos \theta_{lab} < 0.92$ and they are required to originate close to the
beam-beam interaction point. The individual photon energy in the laboratory frame
is required to exceed 30 \mev. The selection of semileptonic $B$ decays is
enhanced by requiring $|\vec{p}_{miss}| > 0.5 \gevc$,
and that $\vec{p}_{miss}$ points into the detector fiducial volume,
$|\cos\theta_{miss}|< 0.9$, thereby effectively reducing the impact
of particle losses close to the beams. Furthermore,
since in semileptonic $B$ decays with a high-momentum electron, the neutrino
and the electron are emitted preferentially in opposite directions, we require
that the angle $\Delta \alpha$ between these two particles fulfill the condition
$\cos \Delta \alpha <0.4$. These requirements for the missing momentum reduce
the continuum background from QED processes and $e^+e^- \ra q \bar{q}$
production by an additional factor of 3, while the signal
loss is less than 20\%.

\begin{figure}[!htb]
\begin{center}
\includegraphics[height=7.0cm]{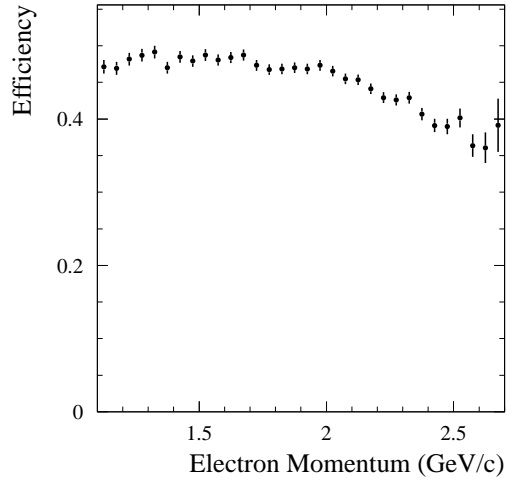}
\caption{
Selection efficiency for events with $B \rightarrow X_u e \nu$ decays as a function
of the electron momentum. The error bars represent the statistical errors.
}
\label{fig:p0}
\end{center}
\end{figure}

The stated selection criteria result in an efficiency (including effects of
bremsstrahlung) of $35 - 50$\% for selecting $B \ra X_u e \nu$ decays; 
its dependence on the electron momentum is shown in Fig.~\ref{fig:p0}.

\subsection{Background Subtraction}
\label{sec:background}

\begin{figure}[!htb]
\begin{center}
\includegraphics[height=8.5cm]{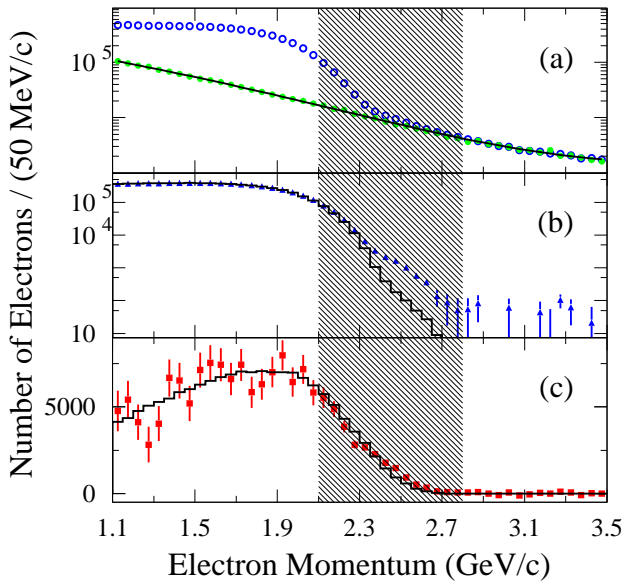}
\caption{
(color online) Electron momentum spectra in the $\FourS$
rest frame: (a) on-resonance data (open circles -- blue),
scaled off-resonance data (solid circles -- green);
the solid line shows the result of the fit
to the non-\BB\ events
using both on- and off-resonance data;
(b) on-resonance data after subtraction of the fitted
non-\BB\ background (triangles -- blue) compared to simulated \BB\ background
that is adjusted by the combined fit to the on- and off-resonance
data (histogram);
(c) on-resonance data after subtraction of all backgrounds
(linear vertical scale, data points -- red), compared to the simulated
$B \ra X_u e \nu$ signal spectrum (histogram);
the error bars indicate errors from the fit, which include the uncertainties
in the fitted scale factors for non-\BB\ and $X_c e \nu$ backgrounds.
The shaded area indicates the momentum interval for which the on-resonance
data are combined into a single bin for the purpose of reducing the
sensitivity of the fit to the shape of the signal spectrum in this region.
}
\label{fig:p1}
\end{center}
\end{figure}

The spectrum of the highest momentum electron in events selected
by the criteria described above is shown in
Fig.~\ref{fig:p1}a, separately for data recorded on and
below the \FourS\ resonance.
The data collected on the $\Upsilon(4S)$ resonance include
contributions from \BB\ events and non-\BB\ background.
The latter is measured using off-resonance data, 
collected below \BB\ production threshold, and using on-resonance
data above 2.8~\gevc, {\it i.e.,} above the endpoint for electrons
from $B$ decays. The \BB\ background to the
$B \to X_u e \nu$ spectrum is estimated
from MC simulation,
with the normalization of the individual contributions determined
by a fit to the total observed spectrum.

\subsubsection{Non-\BB\ Background}

To determine the non-\BB\ background we perform a $\chi^2$ fit to
the off-resonance data in the momentum interval of 1.1
to 3.5~\gevc and to
on-resonance data in the momentum interval of 2.8
to 3.5~\gevc. Since the c.m.\ energy for the off-resonance data is
0.4\% lower than for the on-resonance data, we scale the electron momenta
for the off-resonance data by the ratio of the c.m.\ energies.

The relative normalization for the two data sets is
$$ r_{L} = \frac{s_{\mathit{OFF}}}
{s_{\mathit{ON}}} \frac{\int\! L_{\mathit{ON}}\,dt}
{\int\! L_{\mathit{OFF}}\,dt} = 8.433\pm 0.004\pm 0.021, $$
where $s$ and $\int\! L\,dt$ refer to the c.m.\ energy squared and integrated
luminosity of the two data sets.
The statistical uncertainty of $r_L$ is determined by the number of detected
$\mu^+\mu^-$ pairs used for the measurement of the integrated luminosity; the 
systematic error of the ratio is estimated to be $0.25\%$.

The $\chi^2$ for the fit to the non-\BB\ events is defined as follows,

\begin{equation}
\chi^2_c = \sum_i \frac{(f(\vec{a},p_i)-r_{L} n_i)^2}{r_{L}^{2} n_i}+
\sum_{j
} \frac{(f(\vec{a},p_j)-N_j)^2}{ N_j}.
\end{equation}
\noindent Here $n_i$ and $N_j$ refer to the number of selected events
in the off- and on-resonance samples, for
the $i$-th or $j$-th momentum bin ($p_j > 2.8 \gevc$), and
$\vec{a}$ is the set of free parameters of the fit.
For the function approximating the
momentum spectrum, we have chosen an exponential expression of the form
\begin{equation}
\label{f1}
f(\vec{a},p) = a_1 + \exp(a_2 + a_3 p + a_4 p^2 ) \,.
\end{equation}
\noindent
The fit describes the data well: $\chi^2=70$ for 58 degrees of freedom.
Above 2.8 \gevc, we observe $(36.7 \pm 0.2)\times 10^3$ events in the
on-resonance data, compared to the fitted number of $(36.6 \pm 0.2)\times 10^3$ events.

\subsubsection{\BB\ Background}
The electron spectrum from $B$-meson decays is composed
of several contributions, dominated by
the various semileptonic decays. Hadronic 
$B$ decays contribute mostly via hadron misidentification and
secondary electrons from decays of $D$, $J/\psi$, and $\psi(2S)$
mesons.

We estimate the total background by fitting the observed inclusive electron
spectrum to the sum of the signal and individual background contributions.
For the individual signal and \BB\ background contributions,
we use the MC simulated spectra, and treat their
relative normalization factors as free parameters in the fit.
The non-\BB\ background is parameterized by the exponential functions
$f(\vec{a},p_i)$, as described above.
We expand the $\chi^2$ definition as follows,
\begin{eqnarray}
\chi^2 &=& \sum_i \frac{(f(\vec{a},p_i)-r_{L} n_i)^2}{r_{L}^2 n_i} + \nonumber \\
      & &  \sum_j \frac{(f(\vec{a},p_j)+S(\vec{b},p_j)-N_j)^2}{N_j +
            \sigma^2_{j\,MC}},
\end{eqnarray}
\noindent where the first sum is for the off-resonance data
and the second sum for the on-resonance data. The \BB\ electron
spectrum is approximated as
$S(\vec{b},p_j)=\sum_k b_k g_k(p_j)$, where the free parameters
$b_k$ are the correction factors to the MC default branching fractions for
the six individual contributions $g_k(p_j)$ representing the signal
$B\to X_ue\nu$ decays, the background $B\ra De\nu$, $B\ra D^*e\nu$,
$B\ra D^{**}e\nu$, $B\ra D^{(*)}\pi e\nu$ decays,
and the sum of other background events with electrons from secondary
decays or misidentified hadrons.
$\sigma_{j\,MC}$ is the statistical error of the number of simulated events
in the $j$-th bin. The momentum spectra $g_k(p_j)$ are histograms taken from MC simulations.

\begin{table*}[floatfix]
\caption{
Summary of the signal extraction: the number of events (in units of $10^3$)
for the total sample, the principal background
contributions, and the remaining signal, as well as the signal efficiencies,
for five intervals of the electron momentum.
The errors listed are statistical, including the uncertainties in the fitted scale factors
for the non-\BB\ and $X_c e \nu$ backgrounds. The error values of 0.00 represent errors of
less than 0.005.
}
\label{table:r1}
{\small
\begin{ruledtabular}
\begin{tabular} {lrrrrr}
$\Delta p$ (\gevc)  & 2.0~--~2.6~ & 2.1~--~2.6~ & 2.2~--~2.6~ & 2.3~--~2.6~ & 2.4~--~2.6~  \\ \hline
Total sample    & 609.81 $\pm$ 0.78 &295.76 $\pm$ 0.54 & 133.59 $\pm$ 0.37 & 65.48 $\pm$ 0.26 & 35.38 $\pm$ 0.19 \\ \hline
Non-\BB\ background       & 142.38 $\pm$ 0.63 &105.20 $\pm$ 0.48 &  74.86 $\pm$ 0.36 & 50.13 $\pm$ 0.25 & 29.96 $\pm$ 0.16 \\
$X_c e \nu$ background    & 416.22 $\pm$ 2.52 &157.17 $\pm$ 1.29 &  38.82 $\pm$ 0.47 &  4.00 $\pm$ 0.10 &  0.09 $\pm$ 0.01 \\
$J/\psi$ and $\psi(2S)$
              &   6.17 $\pm$ 0.14 &  4.00 $\pm$ 0.10 &   2.33 $\pm$ 0.06 &  1.17 $\pm$ 0.04 &  0.47 $\pm$ 0.02 \\
Other $e^{\pm}$ background  &   1.61 $\pm$ 0.05 &  0.62 $\pm$ 0.02 &   0.24 $\pm$ 0.01 &  0.08 $\pm$ 0.01 &  0.03 $\pm$ 0.00 \\
$\pi$ mis-identification    &   1.34 $\pm$ 0.04 &  0.98 $\pm$ 0.03 &   0.64 $\pm$ 0.02 &  0.34 $\pm$ 0.02 &  0.10 $\pm$ 0.01 \\
$K$ mis-identification      &   0.47 $\pm$ 0.02 &  0.26 $\pm$ 0.01 &   0.13 $\pm$ 0.01 &  0.05 $\pm$ 0.01 &  0.01 $\pm$ 0.00 \\
Other mis-identification    &   0.27 $\pm$ 0.01 &  0.15 $\pm$ 0.01 &   0.08 $\pm$ 0.01 &  0.04 $\pm$ 0.01 &  0.02 $\pm$ 0.00 \\
$X_u e \nu$ background &  1.62 $\pm$ 0.10 &  0.66 $\pm$ 0.05 &   0.20 $\pm$ 0.02 &  0.03 $\pm$ 0.01 &  0.01 $\pm$ 0.00 \\ \hline
$X_u e\nu$ signal& 39.72 $\pm$ 2.70 & 26.72 $\pm$ 1.49 &  16.31 $\pm$ 0.71 &  9.64 $\pm$ 0.38 &  4.70 $\pm$ 0.25 \\
\hline
$X_u e \nu$ efficiency (\%) 
                 &  42.1 $\pm$ 0.3~        &  41.2 $\pm$ 0.4~  &   40.2 $\pm$ 0.5~ &  39.5 $\pm$  0.7~ & 37.9  $\pm$ 1.0~ \\

\end{tabular}
\end{ruledtabular}
}
\end{table*}

\subsubsection{Fit to Inclusive Spectra}
The fit is performed simultaneously to the on- and off-resonance
electron momentum spectra in the range from 1.1
to 3.5 \gevc, in bins of 50\mevc. The lower part of the spectrum determines
the relative normalization of the various background contributions, allowing
for an extrapolation into the endpoint region above 2.0 \gevc.
To reduce a potential systematic bias from the assumed shape of the signal
spectrum, we combine the on-resonance data for the interval from
2.1 to 2.8 \gevc into a single bin. The lower limit of this bin is chosen
so as to retain the sensitivity to the steeply falling \BB\ background
distributions, while containing a large fraction of the signal events
in a region where the background is low. The fit results are insensitive
to changes in this lower limit in the range of 2.0 to 2.2 \gevc.
The number of signal events in a given momentum interval is taken as the
excess of events above the fitted background.

The observed spectra, the fitted non-\BB\ and
\BB\ backgrounds and the signal are shown and compared to
MC simulations in Fig.~\ref{fig:p1}.
The fit has a $\chi^2$ of $96$ for 73 degrees of freedom.
Above 2.3 \gevc, the non-\BB\ background
is dominant, while at low momenta the semileptonic \BB\ background
dominates. Contributions from hadron misidentification are small,
varying from 6\% to 4\% as the electron momentum increases.
The theoretical prediction for the signal $B \ra X_u e \nu$ spectrum
based on the BLNP calculations uses SF parameters extracted from
the combined fit~\cite{babarcf} to the moments measured by the
\babar\ Collaboration.

The fitting procedure was chosen in recognition of the fact that currently
the branching fractions for the individual $B\ra X_c \ell \nu$
decays are not well enough measured to perform an adequate background subtraction.
The MC simulation takes into account the form factor and angular distributions
for the $B\ra D e \nu$ and $B\ra D^* e \nu$ decays. For decays to higher-mass
mesons, this information is not available. As a result, we do not consider this
fit as a viable method of measuring these individual branching fractions.
Nevertheless, the fitted branching fractions agree reasonably well with the
measured branching fractions~\cite{pdg2002}. For the decays to
higher-mass states, the ability of the fit to distinguish between decays to
$D^{**} e \nu$ and $D^{*}\pi e \nu$ is limited. The sum of the
two contributions, however, agrees with current measurements~\cite{pdg2002}.

Table~\ref{table:r1} shows a summary of the data, principal backgrounds
and the resulting signal. The errors are statistical, but for
the non-\BB\ and $X_c e \nu$ background
they include the uncertainties of the fitted parameters.
The data are shown for five overlapping signal regions, ranging in
width from 600 to 200 \mevc.
We choose 2.6~\gevc\ as the common upper limit of the signal regions
because at higher momenta the signal contributions are very small
compared to the non-\BB\ background. As the lower limit is extended
to 2.0~\gevc , the error on the \BB\ background subtraction increases.

\section{Systematic Errors}
\label{systematics}

A summary of the systematic errors
is given in Table~\ref{table:t2} for five intervals in the electron momentum.
The principal systematic errors originate from the event selection and
the background subtraction. The uncertainty in the event simulation and its
impact on the momentum dependence of the efficiencies for signal and
background are the experimental limitations of the current analysis.
The second largest source of uncertainties is the estimate of the \BB\ background
derived from the fit to the observed electron spectrum, primarily due to the
uncertainties in the simulated momentum spectra of the various contributions.
In addition, there are relatively small corrections to the
momentum spectra due to variations in the beam energies, and radiative
effects.

\begin{table}[floatfix]
\caption{
Summary of the relative systematic errors (\%) on the partial branching fraction
measurements for $B \ra X_u e \nu$ decays, as a function of $p^{min}$, the lower
limit of the signal momentum range. The common upper limit is 2.6 \gevc .
The sensitivity of the signal extraction to the uncertainties in the SF
parameters is listed as an additional systematic error, separately for the
four sets of SF parameters.
}
\label{table:t2}
{\small
\begin{ruledtabular}
\begin{tabular}{lccccc}
$p^{min} (\gevc)$ &
  $2.0 $  &
   $2.1 $  &
   $2.2 $  &
   $2.3 $  & 
   $2.4 $ \\
\hline
Track finding efficiency                    & $0.7$ & $0.7$ & $0.7$ & $0.7$ & $0.7$\\
Electron identification                     & $1.4$ & $1.4$ & $1.3$ & $0.9$ & $0.8$\\
Event selection efficiency                  & $6.8$ & $6.7$ & $6.1$ & $5.5$ & $7.9$\\
Non-\BB\ background                         & $2.4$ & $2.5$ & $2.4$ & $2.5$ & $2.3$\\
$J/\psi$ and $\psi(2S)$ background          & $0.9$ & $0.8$ & $0.8$ & $0.6$ & $0.5$\\
$B \to D^* l\nu$ form factor                & $2.4$ & $2.3$ & $2.0$ & $1.3$ & $0.5$\\
$B \to D l\nu$ form factor                  & $0.7$ & $0.9$ & $0.8$ & $0.2$ & $0.4$\\
$B \to D^{**} e \nu$ spectrum               & $2.8$ & $2.5$ & $2.4$ & $0.9$ & $0.7$\\
Other $e^{\pm}$ background                  & $0.5$ & $0.3$ & $0.2$ & $0.1$ & $0.1$\\
$B \to X_u e \nu$ background                & $1.1$ & $0.6$ & $0.3$ & $0.1$ & $0.0$\\
$\pi$ mis-identification background         & $0.8$ & $0.9$ & $0.9$ & $0.8$ & $0.5$\\
$K$ mis--identification background          & $0.4$ & $0.4$ & $0.3$ & $0.2$ & $0.1$\\
Other hadron mis-identification             & $0.2$ & $0.2$ & $0.2$ & $0.1$ & $0.1$\\
$B$ movement                                & $1.3$ & $1.7$ & $1.5$ & $0.6$ & $0.1$\\
Bremsstrahlung and FSR                      & $1.0$ & $1.2$ & $1.2$ & $0.9$ & $0.9$\\
$N_{B\bar{B}}$ normalization                & $1.1$ & $1.1$ & $1.1$ & $1.1$ & $1.1$\\
\hline
Total experimental error                     & $8.8$ & $8.6$ & $7.9$ & $6.6$ & $8.5$\\
\hline
$ B \ra X_u e\nu$ spectrum                  & & & & & \\
$X_s\gamma $\,SF, fit to spectrum           & $6.0$ & $3.5$ & $1.6$ & $0.3$ & $0.1$\\
$X_s\gamma $\,SF, fit to moments            & $11.3$ & $6.7$ & $3.1$ & $0.6$ & $0.1$\\
$X_c e \nu $\,SF, fit to moments            & $13.3$ & $8.6$ & $4.0$ & $0.8$ & $0.0$\\
SF, combined fit to moments                 & $7.2$ & $4.8$ & $2.3$ & $0.5$ & $0.0$\\
\hline
Total systematic error                      & & & & & \\
$X_s\gamma $\,SF, fit to spectrum           & $10.7$ & $9.3$ & $8.1$ & $6.6$ & $8.5$\\
$X_s\gamma $\,SF, fit to moments            & $14.3$ & $10.9$ & $8.5$ & $6.6$ &$8.5$\\
$X_c e \nu $\,SF, fit to moments            & $15.9$ & $12.2$ & $8.9$ & $6.6$ &$8.5$\\
SF, combined fit to moments                 & $11.4$ & $9.8$ & $8.2$ & $6.6$ & $8.5$\\
\end{tabular}
\end{ruledtabular}
}
\end{table}

\subsection{Detection and Simulation of \boldmath{$B \to X_u e \nu$} Decays}

The selection efficiency for
$B \rightarrow X_u e \nu$ decays is determined by MC simulation.
We include in the uncertainty of the signal spectrum not only the
uncertainty in the simulation of the detector response,
but also the uncertainty in the simulation of the momentum and angular
distributions of the electron, as well as the hadrons and the neutrino.

\subsubsection{Detector related uncertainties}
For a specific model of the signal decays there are three major factors
that determine the efficiency: the track reconstruction for the electron,
the electron identification, and losses due to the detector acceptance
and event selection.

The uncertainty in the tracking efficiency has been studied in detail
and is estimated to be $\sim 0.7\%$ per track.
The average identification efficiency
for electrons with momentum above 1.0~\gevc is estimated to be on average
92\% ~\cite{thorsten}, based on large control samples of radiative Bhabha
events and two-photon interactions.
In \BB\ events the actual efficiencies are slightly lower due to
higher track and photon multiplicity.
This difference decreases gradually from about 2.5\% at 1.0~\gevc to
less than 0.8\% at 2.0~\gevc and above. We add in quadrature
50\% of this observed
difference to the statistical and systematic errors
from the control samples. We assess the impact of this momentum-dependent
uncertainty on the observed electron spectrum for both signal and
background (see below).

\subsubsection{Uncertainties in the signal spectrum}

The momentum distribution of the signal electrons is not precisely
known because many of the exclusive decay modes that make up the total
inclusive $B \rightarrow X_u e \nu$ decays are still
unobserved or poorly measured due to small event samples, and the form factors
for most of the observed exclusive decay modes are not measured.
To evaluate the sensitivity of the signal efficiency to the decay multiplicity and
the shape of the momentum spectrum,
we independently vary the relative contributions of the different decay modes
by their current experimental uncertainties.
We observe changes in the signal yield of less than 3.0\% for the spectrum above
1.1 \gevc, and less than 1.0\% above 2.3\gevc.

The systematic uncertainties inherent in the modeling of the signal decays to
non-resonant hadronic states and their impact on the signal yield have been
studied by varying the SF parameters. We try four sets of SF parameters, two
derived from the recent analysis of the $B\ra X_s \gamma$ decays~\cite{babar_photons}
based on the semi-inclusive photon spectrum and moments derived from this spectrum,
one derived from moments in inclusive $B\ra X_c \ell \nu$ decays~\cite{babarvcb},
and one from a combined fit~\cite{babarcf} to all moments measured by the \babar\
Collaboration. For each set of SF parameters we calculate the signal momentum
spectrum and repeat the fit to the data. We observe small changes in the fitted
$B\ra X_c \ell \nu$ background which result in changes of the signal yield.
Taking into account the errors and correlations of the measured SF parameters,
we derive the errors listed separately in Table~\ref{table:t2}. The impact is
largest for the signal regions extending to lower momenta, where this becomes
the largest source of systematic error.

Not included in this estimate is the sensitivity of the signal yield
to the event selection criteria, specifically those based on the variables
${\cal R}_2$ and $p_{miss}$. These selection criteria influence not only the
signal, but more so the background distributions. Details are discussed below.

\subsection{Non-\boldmath{\BB\ } Background}

Systematic errors in the subtraction of the non-\BB background
could be introduced by the choice of
the fitting function describing the
electron spectrum and by the uncertainty in the relative normalization of the
on- and off-resonance data.

To assess the uncertainty in the shape of this background we have compared fits with
different parameterizations of the fit function. In addition to the exponential
function described above (Eq.\ 2), we have tried linear combinations of Chebyshev
polynomials up to fifth order. The resulting fits are equally consistent with
the data. The differences in the non-\BB\ background estimates between different
parameterizations are less than 0.5\%. Above 2.8 \gevc\, the number of
observed events in the on-resonance data sample agrees to 0.3\% with the
number of events predicted from the fit to the off-resonance sample.

If the relative normalization is treated as a free fit parameter, its deviation
from the value based on luminosity measurements is less than one standard
deviation.
Thus, we use
the more accurate value based on luminosity measurements.
As a systematic error for the non-\BB\ background we take 0.5\% of this
background contribution, which includes the
errors of the normalization factor and the background shape approximation.

\subsection{\boldmath{$B \to J/\psi X$} Background}

$J/\psi$ decays to $e^+e^-$ pairs are vetoed by a restriction on the di-electron
invariant mass. However, this veto is only about 50\% efficient,
primarily because of acceptance losses.
The remaining, mostly single-electron background is estimated from simulation.
We observe a difference of $(5.0 \pm 2.7)\%$ between
the veto efficiency for electron pairs in data and simulation, and thus assign
a 5\% error to the residual background.
This background amounts to $18\%$ of the signal for $p_{e} > 2.0~\gevc $ and 10\%
for $p_{e} > 2.3~\gevc $ and the resulting uncertainty on the signal branching fraction
is estimated to vary from 0.9\% to 0.5\%.
The background from $\psi(2S) \ra e^+e^-$ decays is significantly smaller,
and thus its uncertainty is negligible.

\subsection{\boldmath{\BB\ } Background}

The shapes of the \BB\ backgrounds are derived from MC simulations.
The branching fractions
for exclusive semileptonic $B\to X_c e \nu$ decays
are currently not precisely known. Thus the electron spectra from inclusive
$B\to X_c e \nu$ decays may differ from those of the simulation. For
this reason, we have introduced scale factors in the fits to the electron
spectrum to adjust the relative normalization of the various contributions.
To test the sensitivity to the shape of the dominant contributions, we have varied
the form factors for decays to $D^* e \nu$ and $D e \nu$, and changed the relative
proportion of contributions from narrow and wide resonances to $D^{**} e \nu$ decays.

For $B\to D e \nu$ and $B \ra D^* e \nu$ decays we use HQET
parameterizations~\cite{CLN,GL} of the form factors.
To study the impact of the uncertainties in the measured
form factors, we reweight the MC-simulated spectrum
for a given decay mode to reproduce the change in the spectrum
due to variations of the form-factor parameters, and
repeat the standard fit to the data.
From the observed changes in the signal yield as a function of the choice of
the form-factor parameters for $D^* e \nu$ decays, we assess the systematic error
on the signal yield by taking into account the measured form-factor parameters,
$\rho^2$, $R_1$, and $R_2$, their errors, and their covariance matrix~\cite{babarff}.
For $D e \nu$ decays, we rely on a measurement of $\rho_D^2$ by the CLEO~\cite{bd_form} and
Belle Collaborations~\cite{bd_form2}.
Similarly, we estimate the impact of the uncertainty in $\rho_D^2$ by comparing the
default fit results with spectra corresponding to variations of $\rho_D^2$ by
one standard deviation. We take the shift of the signal yield as a systematic error.

To assess the impact of the poorly known branching fractions for various
$D^{**} e \nu $ decay modes on the shape of the electron spectrum,
we have repeated the fits with the relative branching fractions for the individual
decay modes changed by up to 50\%. As long as we do not eliminate the decays to
the two narrow resonances, $D_1(2437)$ and $D_2(2459)$, we obtain consistent results.
Specifically, if we eliminate the decays involving
the two wider resonances, $D_0(2308)$ and $D_1'(2460)$, the results
change by less than 3\%. We adopt this change as the estimate of
the systematic error due to the uncertainty of decays to $D^{**}$ states.

Similarly, we vary the branching fractions for secondary electrons
from semileptonic $D$ decays by 10\% and adopt the observed change
as a systematic error.
In addition, there is a small contribution from events which contain a
$B\to X_u e \nu$ decay, but contribute to the background rather than the signal,
because the track identified as a signal electron does not originate from
this decay. We estimate the uncertainty of this very small contribution to be 30\%.

For background from hadronic $B$ decays, the uncertainty in the spectrum
is primarily due to the uncertainty in the momentum-dependent hadron
misidentification. The uncertainties of misidentification
probabilities are estimated to be 20\% and 30\%,
for pions and kaons, respectively. The uncertainty
in the fractions of pions and kaons is taken as the difference between
simulated and observed charged particle spectra, which is
about 5\% for pions and kaons.
With these uncertainties in the hadron misidentification backgrounds,
the fractional error in the number of subtracted background events
is $\sim 20$\% for pions and $\sim 30$\% for kaons.
In addition, there is a small background from
protons and from unidentified particles; its total uncertainty is estimated to be
about 50\% smaller than for identified kaons.

\subsection{\label{AB_syst_study}
{Uncertainty in the \boldmath{$B$} Meson Momentum Spectrum}}

The non-zero momentum of the $B$ meson in the \FourS\ rest
frame affects the shape of the electron spectrum near the endpoint.
To estimate the systematic error
associated with the uncertainty in the initial $B$-meson momentum spectrum,
we compare the simulated and measured energy spectra for fully
reconstructed charged $B$ mesons for different data taking periods.
The widths of the energy distributions
agree well for all data, but in some of the data sets we observed a shift in the
central value of up to 2.2 \mev relative to the simulation, which assumes a fixed
center-of-mass energy. We correct
the simulation for the observed shifts, and assess the effect of the uncertainty of
0.13 \mev\ in this shift on the branching fraction measurement.

\subsection{Bremsstrahlung and Radiative Corrections}
\label{Radiative}

For comparison with other experiments and with theoretical calculations, the signal
spectrum resulting from the fit is corrected for bremsstrahlung in the detector and
for final-state radiation. Corrections for QED radiation in the decay process are
simulated using PHOTOS~\cite{photos}. This simulation includes multiple-photon emission
from the electron, but does not include electroweak corrections for quarks. The
accuracy of this simulation has been compared to analytical
calculations performed to ${\cal O}(\alpha)$~\cite{photos}. Based on this comparison
we assign an uncertainty of 20\% to the PHOTOS correction, leading to
an uncertainty in the signal yield of about 1\%.

The uncertainty in the energy loss of electrons due to bremsstrahlung in the beam
pipe and tracking system is determined by the
uncertainty in the thickness of the detector material,
estimated to be
$(0.0450 \pm 0.0014) X_0$ at normal incidence. The thickness of the material
was verified using electrons from Bhabha scattering as a function of the polar
angle relative to the beam. The impact of the uncertainty in the energy loss on
the signal rate was estimated by calculating the impact of an additional
$ 0.0014 X_0$ of material.

\subsection{\label{SES}
{Sensitivity to the Event Selection}}

We have checked the sensitivity of the fits to the electron spectrum to changes in the
event selection. We have also assessed the impact of the momentum-dependent
uncertainty in the electron efficiency on the fitted signal yield.
These variations of the event selection change the signal efficiency and lead to
variations of up to $50\%$ in the size of the non-\BB\ background,
and up to $20\%$ in the $\BB$ background.

Though some of the observed changes in the efficiency-corrected signal yield may already
be covered by the form-factor and other variations, we conclude that these tests do reveal
significant changes that have to be accounted for.

The largest variation (5\%) is observed for changes in the restriction on ratio
of the Fox-Wolfram moments, ${\cal R}_2$, from the default value of 0.5 to 0.6.
Other sizable variations are observed for changes in the restrictions on the absolute
value and direction of the missing momentum vector.
${\cal R}_2$ and the missing momentum are quantities that are derived from the
measured momenta of all charged 
and neutral particles in the event, and are therefore sensitive to even small
differences in data and simulation.
We interpret the observed changes as representative for the uncertainties in the
MC simulation of the selection of signal and background events and adopt the
observed changes between the default fits and the fits with looser selection
criteria as systematic errors. Adding the observed changes in quadrature leads
to a relative systematic error of between 5\% and 8\% on the partial branching fraction.

\section{Results}
\label{results}

\subsection{Determination of the Partial \boldmath{$B\to X_u e \nu$}
Branching Fraction}

For a given interval $\Delta p$ in the electron momentum,
we calculate the inclusive
partial branching fraction $B\to X_u e \nu$ according to
\begin{equation}
\Delta {\cal B} (\Delta p) = \frac{N_{tot}(\Delta p)-N_{bg} (\Delta p) }
     {2\epsilon(\Delta p) N_{\BB}} (1+\delta_{\mathit{rad}}(\Delta p)).
\end{equation}

\noindent
Here $N_{tot}$ refers to the total number of
electron candidates selected in the on-resonance data and $N_{bg}$ refers to
the total background, from non-\BB\ and \BB\ events,
as determined from the fit to the spectrum.
$\epsilon(\Delta p)$ is the total efficiency
for selecting a signal electron from
$B \rightarrow X_u e\nu$ decays (including bremsstrahlung in the detector
material), and $\delta_{\mathit{rad}}$ accounts for the
distortion of the electron spectrum due to final-state radiation.
This is a momentum-dependent correction, derived from the
MC simulation based on PHOTOS~\cite{photos}.
The total number of produced \BB\ events
is $N_{\BB} = (88.36\pm 0.02_{stat} \pm 0.97_{syst}) \times 10^6$.

\begin{figure}[!htb]
\begin{center}
\includegraphics[height=7.5cm]{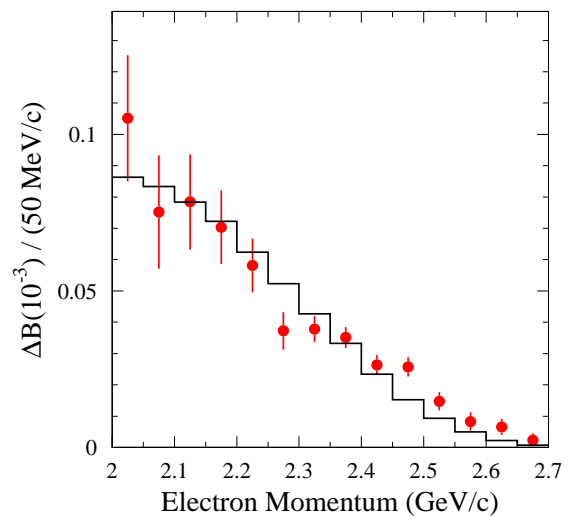}
\caption{
The differential branching fraction for charmless semileptonic $B$ decays
(data points)
as a function of the electron momentum (in the $\FourS$
rest frame) after background subtraction and corrections for bremsstrahlung and
final state radiation, compared to the Monte Carlo simulation (histogram).
The errors indicate the statistical errors on the background subtraction,
including the uncertainties of the fit parameters.
For the signal simulation, the SF parameters are extracted from a combined fit to all \babar\ moments.
}
\label{fig:p2}
\end{center}
\end{figure}

The differential branching fraction as a function of the electron momentum
in the \FourS\ rest frame is shown in
Fig.~\ref{fig:p2}, fully corrected for efficiencies and radiative effects.
The data are well reproduced by the signal simulations using the SF parameters
derived from the combined fit to all moments measured by the \babar\ Collaboration
~\cite{babarcf}, specifically $m_b^{SF}(1.5\,\mathrm{GeV}) = 4.59\,\mathrm{GeV}/c^2$,
$\mu^{2\,SF}_{\pi}(1.5\,\mathrm{GeV}) = 0.21\,\mathrm{GeV}^2$.
The partial branching fractions
for the five overlapping electron momentum intervals are
summarized in Table~\ref{table:br2f_bbrxsgDFN}.
The stated errors on ${\Delta\cal B}$ represent the statistical and total
systematic uncertainties of the measurement, including the uncertainty due
to the sensitivity to the SF parameters, as stated in Table II.
As the lower limit on the electron momentum decreases, the statistical and
systematic errors are more and more dominated by the $B \rightarrow X_c e\nu$
background subtraction.

\subsection{Extraction of the Total Charmless Branching Fraction
and \boldmath{$|V_{ub}|$}}

As mentioned earlier, we use two sets of theoretical calculations
to extract $|V_{ub}|$ from the partial electron spectrum.
The first, and so far the most commonly used, method
derives $|V_{ub}|$ from the total
charmless semileptonic branching fraction
and the average $B$ lifetime, $\tau_b=(1.604 \pm 0.012)\,\mbox{ps}$
~\cite{pdg2004}, as follows,
\begin{eqnarray}
|V_{ub}| &=& 0.00424 \left( \frac{\textrm{\BR}(B \to X_u l \nu)}{0.002}
\frac{1.61\,\mbox{ps}} {\tau_b} \right)^{1/2} \nonumber \\
        &\times& (1.0 \pm 0.028_{\mathit{pert+nonpert}} \pm 0.039_{m_b}).
\end{eqnarray}
\noindent
Here the first error represents the linear sum of the
uncertainties of the perturbative and non-perturbative
QCD corrections, and the second error is due to the uncertainty in $m_b$. An
overall correction of 0.7\% is included to account for QED corrections.
This formulation~\cite{vub1,vub2,vub3} has been updated to
take into account the recent measurement~\cite{babarvcb}
of $m_b$, $\mu_{\pi}^2$, and other parameters of the heavy quark expansion
in the kinetic mass scheme, specifically
$m_b^{kin}(1.0 \gev)=(4.61 \pm 0.07) \gevcc$ and
$\mu_{\pi}^{2(kin)}(1.0 \gev) = (0.45\pm 0.05) \gev^2$.

We determine the total branching fraction,
\begin{equation}
{\cal B}(B\rightarrow X_u e\nu) = \Delta{\cal B}(\Delta p)
/f_u(\Delta p),
\end{equation}
where $f_u(\Delta p)$ is the fraction of the electron spectrum in a given momentum
interval $\Delta p$. The values of $f_u(\Delta p)$
are estimated based on the DN calculations~\cite{dFN} using the exponential
parameterization of the SF, with the SF parameters
extracted from fits to the
photon spectrum in semi-inclusive $B\ra X_s \gamma$ decays, as measured by the
\babar\ Collaboration~\cite{babar_photons},
$\bar{\Lambda}^{SF}=(0.49^{+0.10}_{-0.06})\gevcc$,
$\lambda_1^{SF} =( -0.24^{+0.09}_{-0.18}) \gev^2$, with a correlation coefficient of $-0.94$.
We obtain very similar results for the two other functional forms suggested to describe the SF~\cite{dFN}.

The results for the predicted fraction $f_u(\Delta p)$, the total charmless
branching fraction, $\cal B$, and $|V_{ub}|$ are presented in Table~\ref{table:br2f_bbrxsgDFN}.
The first error on $f_u$ refers
to the experimental error of the SF parameters from the measurement of the inclusive
photon spectrum. It includes the uncertainty of the background subtraction and
the extrapolation to decays to unmeasured $X_s$ states.
We have taken into account the stated error of the SF parameters, including their correlation.
Specifically, we have taken as an error on $f_u(\Delta p)$ the maximum deviation of the
$f_u(\Delta p)$ from its central value for selected values of the SF
parameters on the error ellipse. The second error accounts for the
dependence on the $\alpha_s$ scale, for the uncertainty in the form of the SF, and for
the uncertainty in the theoretical prediction of $f_u$ from
the $B\to X_s\gamma$ measurement. As suggested by M.\ Neubert \cite{private}
this error has been estimated by varying the
values of $\bar{\Lambda}^{SF}$ and $\lambda_1^{SF}$ by 10\%.

\begin{table*}[floatfix]
\caption{
The partial ($\Delta {\cal B}$) and total (${\cal B}$) branching
fraction for inclusive $B \to X_u e \nu$ decays and $|V_{ub}|$
for five electron momentum intervals. The spectral fractions $f_u$ are
determined using SF parameters extracted from a fit to the
photon spectrum in $B \to X_s \gamma$ decays~\cite{babar_photons}
based on calculations by DeFazio and Neubert~\cite{dFN} and Kagan
and Neubert~\cite{kagan_neubert}. The errors
are explained in the text.
}
\label{table:br2f_bbrxsgDFN}
{\small
\begin{ruledtabular}
\begin{tabular}{ccccc}
$\Delta p $ (\gevc) & $\Delta {\cal B}\,(10^{-3})$ & $f_u(\Delta p)$ &
${\cal B}\,(10^{-3})$ & $|V_{ub}|\,(10^{-3})$ \\ \hline
$2.0 - 2.6$ & $0.479\pm 0.033\pm 0.050$ & $0.298\pm 0.029\pm 0.015$ &
 $1.61\pm 0.18\pm 0.25\pm 0.08$ & $3.80\pm 0.21\pm 0.29\pm 0.10\pm 0.18$ \\
$2.1 - 2.6$ & $0.350\pm 0.020\pm 0.033$ & $0.222\pm 0.026\pm 0.016$ &
 $1.58\pm 0.16\pm 0.24\pm 0.11$ & $3.77\pm 0.19\pm 0.28\pm 0.13\pm 0.18$ \\
$2.2 - 2.6$ & $0.231\pm 0.010\pm 0.018$ & $0.149\pm 0.020\pm 0.016$ &
 $1.55\pm 0.14\pm 0.23\pm 0.16$ & $3.73\pm 0.17\pm 0.27\pm 0.19\pm 0.18$ \\
$2.3 - 2.6$ & $0.146\pm 0.006\pm 0.010$ & $0.086\pm 0.013\pm 0.013$ &
 $1.71\pm 0.13\pm 0.25\pm 0.26$ & $3.92\pm 0.15\pm 0.29\pm 0.30\pm 0.19$ \\
$2.4 - 2.6$ & $0.075\pm 0.004\pm 0.006$ & $0.039\pm 0.006\pm 0.009$ &
 $1.95\pm 0.20\pm 0.32\pm 0.45$ & $4.18\pm 0.21\pm 0.35\pm 0.48\pm 0.20$ \\
\end{tabular}
\end{ruledtabular}
}
\end{table*}

The errors listed for $\cal B$ and $|V_{ub}|$ are specified as follows.
The first error reflects the error on the measurement of
$\Delta {\cal B}$, which includes statistical and experimental systematic
uncertainties, except for the uncertainty in the SF parameters.
The second error is due to experimental uncertainty of SF parameters affecting
both $f_u(\Delta p)$ and $\Delta {\cal B}$.
The third error is the theoretical uncertainty of $f_u(\Delta p)$.
The fourth error on $|V_{ub}|$ accounts for the theoretical
uncertainty in the translation from $\cal B$ to $|V_{ub}|$,
as specified in Eq.\ 5. This error also depends on the $b$-quark mass and thus
is correlated with the theoretical uncertainty on the SF.

The results for the total branching fraction ${\cal B}$ and $|V_{ub}|$ obtained from
the different momentum intervals are consistent within the experimental and theoretical
uncertainties. For intervals extending below 2.3\gevc, the total errors on
$\cal B$ and $|V_{ub}|$ do not depend very strongly on the chosen momentum interval.
While the errors on $\Delta {\cal B}$ are smallest above the kinematic endpoint
for $B\ra X_c e \nu$ decays, the dominant uncertainty arises from
the determination of the fraction $f_u$ and increases substantially
with higher momentum cut-offs. The stated theoretical errors on $f_u$,
acknowledged as being underestimated~\cite{private},
do not include uncertainties from
weak annihilation and other power-suppressed corrections.
Assuming that one can combine the experimental and theoretical errors in quadrature,
the best measurement of the total branching fraction is obtained for the momentum
interval $2.0\,$--$\,2.6$~\gevc.

Though the \babar\ measurement of the photon spectrum~\cite{babar_photons} 
results in the best estimate for the SF parameters, we have also 
considered sets of SF parameters obtained from photon
spectra measured by the CLEO~\cite{cleo_photons} and Belle 
~\cite{belle_photons} Collaborations. These parameters are listed in
Table~\ref{table:sf_all}.
In Table~\ref{table:comparison} the results
obtained for these different SF parameters based on the
\babar\ semileptonic data and on the DN calculations are listed for
the momentum interval $2.0 - 2.6 \gevc$.
The differences between the SF parameters obtained by the CLEO and Belle
Collaborations and the \babar\ results are comparable to the experimental
errors on these parameters.
These differences affect the signal spectrum, and thereby the fitted
background yield. The effect is small for high momentum region
and increases for the signal intervals extending to lower momenta.
The impact of the SF parameters on the partial branching fractions is included
in the total error (see Table~\ref{table:t2}).

\begin{table}[floatfix]
\caption{
SF parameters (at a scale of 1.5 \gev ) measured by different
experiments, based on two different theoretical calculations, top:
DN~\cite{dFN,kagan_neubert}, bottom: BLNP~\cite{neubert_extract}.
}
\label{table:sf_all}
{\small
\begin{ruledtabular}
\begin{tabular}{lrll}
\multicolumn{2}{c}{\rule[-0mm]{0mm}{4mm} Experiment \hfill SF Input} & 
$\bar{\Lambda}\,\,$ (\gevcc) 
& $ \lambda_1$ ($\gev^2$) \\ \hline

\rule[-0mm]{0mm}{4mm}
\babar    & (spectrum)~$X_s \gamma$ ~\cite{babar_photons}
& $0.49 ^{+0.10}_{-0.06} $
& $ -0.24^{+0.09}_{-0.18} $  \\

\rule[-0mm]{0mm}{4mm}
CLEO     & (spectrum)~$X_s \gamma$ ~\cite{cleo_photons}
& $ 0.54 ^{+0.26}_{-0.11} $
& $ -0.34^{+0.18}_{-0.88} $  \\

\rule[-0mm]{0mm}{4mm}
Belle    & (spectrum)~$X_s \gamma$ ~\cite{belle_photons}
& $ 0.66 ^{+0.09}_{-0.06} $
& $ -0.40^{+0.17}_{-0.32} $ \\
\hline
\multicolumn{2}{c}{\rule[-0mm]{0mm}{4mm} Experiment \hfill SF Input} & 
$\bar{\Lambda}\,\,$ (\gevcc)
& $ \mu_{\pi}^2$ ($\gev^2$) \\ 
\hline
\rule[-0mm]{0mm}{4mm}
\babar  & (spectrum)~$X_s \gamma$ ~\cite{babar_photons}
& $ 0.61 ^{+0.07}_{-0.07} $
& $ 0.16^{+0.10}_{-0.08} $ \\
\rule[-0mm]{0mm}{4mm}
\babar  & (moments)~$X_s \gamma$ ~\cite{babar_photons}
& $ 0.75 ^{+0.11}_{-0.13} $
& $ 0.35^{+0.11}_{-0.15} $ \\
\rule[-0mm]{0mm}{4mm}
\babar   & (moments)~$X_c \ell \nu$ ~\cite{babarvcb}
& $ 0.67 \pm 0.08 $
& $ 0.15 \pm 0.07 $  \\
\rule[-0mm]{0mm}{4mm}
\babar   & (comb1.\ moments) ~\cite{babarcf}
& $ 0.69 \pm 0.05 $
& $ 0.21 \pm 0.05 $  \\
\end{tabular}
\end{ruledtabular}
}
\end{table}

\begin{table*}[floatfix]
\caption{
Comparison of measurements of the partial ($\Delta {\cal B}$) and total (${\cal B}$)
branching fraction for inclusive $B \to X_u e \nu$ decays and $|V_{ub}|$
for the electron momentum interval 2.0 to 2.6 \gevc.
The results are obtained for SF parameters
(listed in Table IV) extracted from different experiments.
The first three measurements are based on DN~\cite{dFN,kagan_neubert} calculations,
the remaining four on BLNP~\cite{neubert_extract} calculations, based on
SF parameters extracted from the photon spectra and energy moments, the
$B \ra X_c \ell \nu$ moments~\cite{babarvcb}, and a combined fit to moments~\cite{babarcf}.
The errors are explained in the text.
}
\label{table:comparison}
{\small
\begin{ruledtabular}
\begin{tabular}{llccc}
Experiment  & SF input 
& $\Delta {\cal B}\,(10^{-3})$ 
& ${\cal B}\,(10^{-3})$
& $|V_{ub}|\,(10^{-3})$ \\ \hline
\rule[-0mm]{0mm}{4mm}
\babar & $X_s \gamma$ (spectrum)
& $0.479 \pm 0.033 \pm 0.050$
& $1.61 \pm 0.18 \pm 0.25 \pm 0.08 $ & $3.80 \pm 0.21 \pm 0.29 \pm 0.10 \pm 0.18$  \\
\rule[-0mm]{0mm}{4mm}
CLEO  & $X_s \gamma$ (spectrum)
& $0.491 \pm 0.036 \pm 0.061$
& $1.75 \pm 0.20 \pm 0.48 \pm 0.11 $ & $3.97 \pm 0.23 \pm 0.54 \pm 0.12 \pm 0.19$ \\
\rule[-0mm]{0mm}{4mm}
Belle & $X_s \gamma$ (spectrum)
& $0.548 \pm 0.038 \pm 0.057$
& $2.24 \pm 0.25 \pm 0.27 \pm 0.20 $ & $4.48 \pm 0.25 \pm 0.27 \pm 0.20 \pm 0.22$ \\
\hline
\rule[-0mm]{0mm}{4mm}
\babar & $X_s \gamma$ (spectrum)
& $0.514 \pm 0.037\pm 0.055$
& $1.81 \pm 0.20 \, \, \,^{+0.32}_{-0.24}\, \pm 0.11$ & $3.80 \pm 0.21 \, \,^{+0.49}_{-0.39}\, \pm 0.18$ \\
\rule[-0mm]{0mm}{4mm}
\babar & $X_s \gamma$ (moments)
& $0.577 \pm 0.041\pm 0.082$
& $2.57 \pm 0.29 \, \,\,^{+1.16}_{-0.66}\, \pm 0.23$ & $4.86 \pm 0.28 \, \,^{+1.20}_{-0.89}\, \pm 0.26$ \\
\rule[-0mm]{0mm}{4mm}
\babar & $X_c \ell \nu$ (moments)
& $0.569 \pm 0.039\pm 0.090$
& $2.17 \pm 0.24 \, \,\,^{+0.58}_{-0.41}\, \pm 0.15$ & $4.30 \pm 0.24 \, \,^{+0.75}_{-0.59}\, \pm 0.21$ \\
\rule[-0mm]{0mm}{4mm}
\babar & combined fit to moments
& $0.572 \pm 0.041\pm 0.065$
& $2.27 \pm 0.26 \, \,\,^{+0.33}_{-0.28}\, \pm 0.17$ & $4.44 \pm 0.25 \, \,^{+0.42}_{-0.38}\, \pm 0.22$ \\

\end{tabular}
\end{ruledtabular}
}
\end{table*}

\begin{figure}[!htb]
\begin{center}
\includegraphics[height=7.5cm]{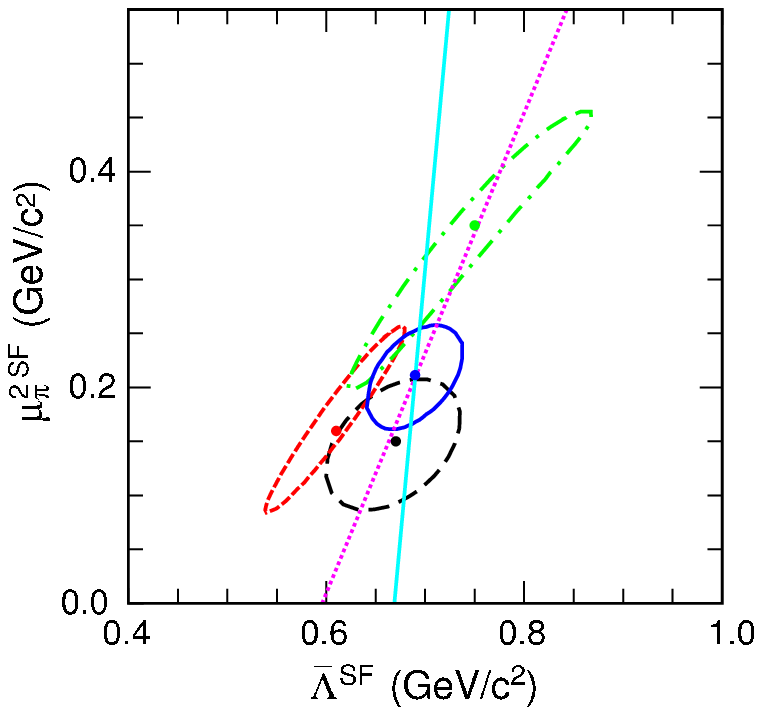}
\caption{
(color online) The fitted values and contours corresponding to
$\Delta \chi^2=1$ for the four sets of SF parameters (see Table IV)
based on the calculations of BLNP, extracted from the photon energy
spectrum (short dash - red) and from the photon energy moments
(dot-dash - green) in $B\ra X_s \gamma$, from the lepton energy and
hadron mass moments in $B\ra X_c e \nu$ decays (long dash - black),
as well as from the combined fit to moments (solid - blue) measured
by the \babar\ Collaboration. Also shown are two straight lines indicting
values of the SF parameters, for which the partial branching fraction
(dotted - magenta) and $|V_{ub}|$ (solid - light blue) are constant.
}
\label{fig:p4}
\end{center}
\end{figure}

The second method for extracting $|V_{ub}|$ is based on recent BLNP
calculations~\cite{neubert_extract}. In this framework the
partial branching fraction $\Delta {\cal B}$ is related directly to $|V_{ub}|$:
\begin{equation}
 |V_{ub}| = \sqrt{\frac {\Delta {\cal B}}{\tau_b \, \zeta(\Delta p)}},
\end{equation}
where $\zeta(\Delta p)$ is the prediction for the partial rate for
$B\ra X_u e \nu$ decays (in units of $\mathrm{ps}^{-1}$).
In these calculations
the leading order SF is constrained by the HQE parameters,
obtained either from the $B\ra X_s \gamma$ or $B\ra X_c e \nu$ decays,
or both.

The values of the SF parameters extracted from
the \babar\ analyses of inclusive $B\to X_s\gamma$~\cite{babar_photons},
$B\to X_c e\nu$~\cite{babarvcb} decays, and the combined fit~\cite{babarcf} to all moments
measured by the \babar\ Collaboration are listed in Table~\ref{table:sf_all}.
Note that the definitions of shape functions and the SF parameters are different for
the DN and BLNP calculations. The different SF parameters and their measurement errors
are also shown in Fig.~\ref{fig:p4}.

The SF parameters based on $B\ra X_s \gamma$ data only are extracted from
either a fit to the photon spectrum or to the first and second moments of
this spectrum in the ``shape function'' scheme.
The HQE parameters extracted from fits to measured moments in the kinetic mass
scheme have been translated into the ``shape function'' scheme at the appropriate scale.
Specifically, the HQE parameters extracted from the moments in $B\ra X_c e \nu$ decays
have been translated based on two-loop calculations~\cite{neubert_loops}.
The HQE parameters resulting from the combined fit to moments of the photon, lepton,
and hadron mass spectra in the kinetic scheme are used to predict the first and second
moments of the photon spectrum down to photon energies of 1.6\gev, based on calculations
by Benson, Bigi, and Uraltsev~\cite{kinetic}. The lower limit on the photon energy is chosen
such that the estimated cut-induced perturbative and non-perturbative corrections to the HQE
are negligible. From these predicted moments, the SF parameters are extracted using
the next-to-leading order calculations in a framework that is consistent with the one
used for the determination of $|V_{ub}|$~\cite{neubert_extract}.

The smallest errors on the SF parameters
are obtained from the
fit to the photon spectrum and the combined fit to all moments.
The fit to the photon spectrum is most sensitive to the high end of the photon
energy spectrum, and relies on the theoretical prediction
for the shape of the spectrum down to low photon energies. Since this shape is
not directly calculable, several forms of the SF are used to assess the
uncertainty of this approach. The use of two sets of the first and second moments
of the photon spectrum, above 1.90 and above 2.09 \gev, is less powerful, due to
much larger statistical and systematic errors, but insensitive to the theoretical
knowledge of the detailed shape of the spectrum.
The SF parameters obtained from moments of the photon spectrum above 1.90 \gevc\
agree with those obtained from the global fit to the moments, but also have larger errors.
Nevertheless, the inclusion of the photon energy moments significantly improves the
sensitivity of the global fit to more than 30 measured moments.

The results for the partial branching fractions $\Delta {\cal B}$ and $|V_{ub}|$
based on the BLNP calculations are listed in
Tables~\ref{table:br2f_bbrxsg_direct}, \ref{table:br2f_bbrxsgm_direct},
\ref{table:br2f_bbrxcenu_direct}, and \ref{table:br2f_cf_direct} for the four
sets of SF parameters.

\begin{table*}[floatfix]
\caption{
The partial branching fraction $\Delta {\cal B}$, the
predicted partial rate $\zeta$ for $B \to X_u e \nu$ decays and $|V_{ub}|$ for
five electron momentum intervals, using the SF parameters
from the photon spectrum in semi-inclusive $B \to X_s \gamma$ decays, 
($\bar{\Lambda}^{SF}= (0.61^{+0.07}_{-0.07}) \,\mathrm{GeV}/c^2$,
$\mu_\pi^{2\,SF}= (0.16^{+0.10}_{-0.08}) \,\mathrm{GeV}^2$, fit to spectrum)~\cite{babar_photons}
based on BLNP calculations~\cite{neubert_extract}.
The errors are explained in the text.
}
\label{table:br2f_bbrxsg_direct}
{\small
\begin{ruledtabular}
\begin{tabular}{ccrc}
$\Delta p $ (\gevc) & $\Delta {\cal B}\,(10^{-3})$ &
$\zeta(\Delta p) (\mathrm{ps}^{-1})$ &
$|V_{ub}|\,(10^{-3})$ \\ \hline
\rule[-0mm]{0mm}{4mm}
$2.0 - 2.6$ &$0.514\pm 0.037\pm 0.055$& $22.2\pm 3.9\pm 2.1$ & $3.80\pm 0.21\, \,\,^{+0.49}_{-0.39}\,\pm 0.18$ \\
\rule[-0mm]{0mm}{4mm}
$2.1 - 2.6$ &$0.366\pm 0.021\pm 0.034$& $16.1\pm 3.3\pm 1.7$ & $3.76\pm 0.20\, \,\,^{+0.49}_{-0.39}\,\pm 0.20$ \\
\rule[-0mm]{0mm}{4mm}
$2.2 - 2.6$ &$0.236\pm 0.011\pm 0.019$& $10.5\pm 2.4\pm 1.5$ & $3.75\pm 0.17\, \,\,^{+0.49}_{-0.40}\,\pm 0.27$ \\
\rule[-0mm]{0mm}{4mm}
$2.3 - 2.6$ &$0.147\pm 0.006\pm 0.010$& $ 5.8\pm 1.5\pm 1.6$ & $3.98\pm 0.16\, \,\,^{+0.53}_{-0.43}\, \pm 0.55$\\
\rule[-0mm]{0mm}{4mm}
$2.4 - 2.6$ &$0.075\pm 0.004\pm 0.006$& $ 2.5\pm 0.7\pm 2.1$ & $4.32\pm 0.22\, \,\,^{+0.59}_{-0.49}\, \pm 1.81$\\
\end{tabular}
\end{ruledtabular}
}
\end{table*}

\begin{table*}[floatfix]
\caption{
The partial branching fraction $\Delta {\cal B}$, the
predicted partial rate $\zeta$ for $B \to X_u e \nu$ decays and $|V_{ub}|$ for
five electron momentum intervals, using the SF parameters
from the moments of the photon energy spectrum in semi-inclusive $B \to X_s \gamma$ decays,
($\bar{\Lambda}^{SF}= (0.75^{+0.11}_{-0.13}) \,\mathrm{GeV}/c^2$,
$\mu_\pi^{2\,SF}= (0.35^{+0.11}_{-0.15}) \,\mathrm{GeV}^2$, fit to
moments)~\cite{babar_photons}
based on BLNP calculations~\cite{neubert_extract}.
The errors are explained in the text.
}
\label{table:br2f_bbrxsgm_direct}
{\small
\begin{ruledtabular}
\begin{tabular}{ccrc}
$\Delta p $ (\gevc) & $\Delta {\cal B}\,(10^{-3})$ &
$\zeta(\Delta p) (\mathrm{ps}^{-1})$ & 
$|V_{ub}|\,(10^{-3})$ \\ \hline
\rule[-0mm]{0mm}{4mm}
$2.0 - 2.6$ &$0.577\pm 0.041\pm 0.082$& $15.2\pm 5.1\pm 1.6$ & $4.86\pm 0.28\, \,\,^{+1.20}_{-0.89}\,\pm 0.26$ \\
\rule[-0mm]{0mm}{4mm}
$2.1 - 2.6$ &$0.392\pm 0.024\pm 0.043$& $10.5\pm 4.1\pm 1.5$ & $4.82\pm 0.25\, \,\,^{+1.20}_{-0.88}\,\pm 0.33$ \\
\rule[-0mm]{0mm}{4mm}
$2.2 - 2.6$ &$0.243\pm 0.011\pm 0.021$& $ 6.5\pm 3.0\pm 1.4$ & $4.82\pm 0.22\, \,\,^{+1.24}_{-0.89}\,\pm 0.52$ \\
\rule[-0mm]{0mm}{4mm}
$2.3 - 2.6$ &$0.148\pm 0.006\pm 0.010$& $ 3.5\pm 1.8\pm 1.6$ & $5.15\pm 0.20\, \,\,^{+1.39}_{-0.98}\,\pm 1.20$\\
\rule[-0mm]{0mm}{4mm}
$2.4 - 2.6$ &$0.075\pm 0.004\pm 0.006$& $ 1.5\pm 0.8 \,\,^{+2.3}_{-1.5}\,$ & $5.62\pm 0.29\, \,\,^{+1.61}_{-1.14}\,\pm 4.27$\\
\end{tabular}
\end{ruledtabular}
}
\end{table*}

\begin{table*}[floatfix]
\caption{
The partial branching fraction $\Delta {\cal B}$, the
predicted partial rate $\zeta$ for $B \to X_u e \nu$ decays and $|V_{ub}|$ for five electron momentum intervals,
based on the SF parameters from
hadron mass and lepton moments in $B \to X_c e \nu$ decays ($\bar{\Lambda}^{SF}= (0.67\pm 0.08) \,\mathrm{GeV}/c^2$,
$\mu_\pi^{2\,SF}= (0.15\pm 0.07) \,\mathrm{GeV}^2$) \cite{babarvcb}
based on BLNP calculations~\cite{neubert_extract}.
The errors are explained in the text.
}
\label{table:br2f_bbrxcenu_direct}
{\small
\begin{ruledtabular}
\begin{tabular}{ccrc}
$\Delta p $ (\gevc) & $\Delta {\cal B}\,(10^{-3})$ &
$\zeta(\Delta p) (\mathrm{ps}^{-1})$ & 
$|V_{ub}|\,(10^{-3}\,$ \\ \hline
\rule[-0mm]{0mm}{4mm}
$2.0 - 2.6$ &$0.569\pm 0.039\pm 0.090$& $19.2\pm 3.9\pm 1.9$ & $4.30\pm 0.24\, \,\,^{+0.75}_{-0.59}\, \pm 0.21$ \\
\rule[-0mm]{0mm}{4mm}
$2.1 - 2.6$ &$0.391\pm 0.022\pm 0.048$& $13.5\pm 3.3\pm 1.6$ & $4.26\pm 0.22\, \,\,^{+0.74}_{-0.58}\, \pm 0.25$ \\
\rule[-0mm]{0mm}{4mm}
$2.2 - 2.6$ &$0.243\pm 0.011\pm 0.022$& $ 8.3\pm 2.6\pm 1.5$ & $4.27\pm 0.19\, \,\,^{+0.77}_{-0.61}\, \pm 0.37$ \\
\rule[-0mm]{0mm}{4mm}
$2.3 - 2.6$ &$0.148\pm 0.006\pm 0.010$& $ 4.3\pm 1.7\pm 1.7$ & $4.65\pm 0.18\, \,\,^{+0.93}_{-0.75}\, \pm 0.91$ \\
\rule[-0mm]{0mm}{4mm}
$2.4 - 2.6$ &$0.075\pm 0.004\pm 0.006$& $ 1.7\pm 0.8 \,\,^{+2.4}_{-1.7}\,$ & $5.28\pm 0.27\, \,\,^{+1.29}_{-1.00}\, \pm 3.76$\\
\end{tabular}
\end{ruledtabular}
}
\end{table*}

\begin{table*}[floatfix]
\caption{
The partial branching fraction $\Delta {\cal B}$, the
predicted partial rate $\zeta$ for $B \to X_u e \nu$ decays and $|V_{ub}|$ for five electron momentum intervals,
based on the SF parameters from the combined fit to \babar\ moments
($\bar{\Lambda}^{SF}= (0.69\pm 0.05) \,\mathrm{GeV}/c^2$,
$\mu_\pi^{2\,SF}= (0.21\pm 0.05)\,\mathrm{GeV}^2$) \cite{babarcf}
based on BLNP calculations~\cite{neubert_extract}.
The errors are explained in the text.
}
\label{table:br2f_cf_direct}
{\small
\begin{ruledtabular}
\begin{tabular}{ccrc}
$\Delta p $ (\gevc) & $\Delta {\cal B}\,(10^{-3})$ &
$\zeta(\Delta p) ( \mathrm{ps}^{-1})$ & 
$|V_{ub}|\,(10^{-3})$ \\ \hline
\rule[-0mm]{0mm}{4mm}
$2.0 - 2.6$ &$0.572\pm 0.041\pm 0.065$& $18.1\pm 2.3\pm 1.8$ & $4.44\pm 0.25\, \,\,^{+0.42}_{-0.38}\, \pm 0.22$ \\
\rule[-0mm]{0mm}{4mm}
$2.1 - 2.6$ &$0.392\pm 0.023\pm 0.038$& $12.6\pm 1.9\pm 1.5$ & $4.40\pm 0.23\, \,\,^{+0.42}_{-0.38}\, \pm 0.27$ \\
\rule[-0mm]{0mm}{4mm}
$2.2 - 2.6$ &$0.243\pm 0.011\pm 0.020$& $ 7.8\pm 1.4\pm 1.4$ & $4.40\pm 0.20\, \,\,^{+0.43}_{-0.39}\, \pm 0.41$ \\
\rule[-0mm]{0mm}{4mm}
$2.3 - 2.6$ &$0.148\pm 0.006\pm 0.010$& $ 4.1\pm 0.9\pm 1.7$ & $4.74\pm 0.18\, \,\,^{+0.52}_{-0.45}\, \pm 0.96$\\
\rule[-0mm]{0mm}{4mm}
$2.4 - 2.6$ &$0.075\pm 0.004\pm 0.006$& $ 1.7\pm 0.5 \,\,^{+2.3}_{-1.7}\,$ & $5.29\pm 0.27\, \,\,^{+0.74}_{-0.59}\,\pm 3.66$\\
\end{tabular}
\end{ruledtabular}
}
\end{table*}

\begin{table}[floatfix]
\caption{
The total (${\cal B}$) branching
fraction for inclusive $B \to X_u e \nu$ decays
for five electron momentum intervals. The spectral fractions $f_u$ are
based on calculations
by Lange, Neubert and Paz~\cite{neubert_extract}
using SF parameters extracted from the combined fit~\cite{babarcf} to
all \babar\ moments. The error definitions are the same as in
Table~III, and they are explained in the text above.
}
\label{table:br2f_cf}
{\small
\begin{ruledtabular}
\begin{tabular}{ccc}
$\Delta p $ (\gevc) & $f_u(\Delta p)$ & ${\cal B}\,(10^{-3})$ \\ \hline
\rule[-0mm]{0mm}{4mm}
$2.0 - 2.6$ & $0.252\pm 0.018\pm 0.019$ & $2.27\pm 0.26\,\,\,^{+0.33}_{-0.28}\,\pm 0.17$ \\
\rule[-0mm]{0mm}{4mm}
$2.1 - 2.6$ & $0.176\pm 0.017\pm 0.019$ & $2.22\pm 0.23\,\,\,^{+0.32}_{-0.27}\,\pm 0.24$ \\
\rule[-0mm]{0mm}{4mm}
$2.2 - 2.6$ & $0.109\pm 0.014\pm 0.020$ & $2.22\pm 0.20\,\,\,^{+0.33}_{-0.28}\,\pm 0.40$ \\
\rule[-0mm]{0mm}{4mm}
$2.3 - 2.6$ & $0.057\pm 0.009\pm 0.023$ & $2.58\pm 0.20\,\,\,^{+0.47}_{-0.36}\,\pm 1.04$ \\
\rule[-0mm]{0mm}{4mm}
$2.4 - 2.6$ & $0.023\pm 0.005\pm 0.033$ & $3.21\pm 0.33\,\,\,^{+0.80}_{-0.58}\,
\,\,^{+4.47}_{-3.21}\, $ \\
\end{tabular}
\end{ruledtabular}
}
\end{table}

The errors cited in these tables are defined and determined in analogy to
those in Table~\ref{table:br2f_bbrxsgDFN}.
The first error on the predicted rate $\zeta$ accounts for the uncertainty due
to the errors in measured parameters
of the leading SF, the second error refers to the theoretical uncertainties in the
subleading SFs, and variations of scale matching, as well as weak annihilation effects.
For $|V_{ub}|$, the first error is the experimental error on the partial branching fraction,
which includes the statistical and the experimental systematic uncertainty,
the second error includes systematic uncertainties on the partial branching fraction
and $\zeta$ due to the uncertainty of the SF
parameters, and the third error is the theoretical uncertainty on $\zeta$,
estimated using the prescription suggested by BLNP.

In Table~\ref{table:comparison} the results
obtained for these different SF parameters based on the
\babar\ semileptonic data and on the BLNP (and DN) calculations are listed for
the momentum interval $2.0 - 2.6 \gevc$. The observed differences are consistent
with the total error stated; they are largest for the SF parameters extracted
from the fit to the photon spectrum as compared to the moments of the photon spectrum.

For all four sets of SFs we observe a tendency for the total
branching fraction, and therefore also $|V_{ub}|$, to be slightly larger at the
higher momentum intervals, but the uncertainties in the predicted rates $\zeta$
are very large for the highest momentum interval.

Based on the BLNP calculations~\cite{neubert_extract} of the inclusive lepton
spectra, we have also determined the total $B\to X_u e\nu$ branching fraction.
The results are presented in Table~\ref{table:br2f_cf}.

The results for $|V_{ub}|$ extracted for the BLNP calculations are close to
those obtained for the DN calculations (see Table~\ref{table:comparison}).
In fact, the results based on the fit to the photon spectrum measured by the
\babar\ Collaboration are identical for all electron momentum ranges, even though
the partial branching fractions differ
by one standard deviation of the experimental error (see Tables III and VI).
Changing the ansatz for the SF from the exponential to a hyperbolic
function~\cite{neubert_extract} has no impact on the results.

\section{Conclusions}

In summary, we have measured the inclusive electron spectrum in charmless
semileptonic $B$ decays and derived partial branching fractions in five overlapping
electron momentum intervals close to the kinematic endpoint. We have extracted the
partial and total branching fractions and the magnitude of the CKM element $|V_{ub}|$
based on two sets of calculations: the earlier ones by DeFazio and Neubert~\cite{dFN}
and Kagan and Neubert~\cite{kagan_neubert}, and the more comprehensive calculations
by Lange, Neubert and Paz~\cite{neubert_extract}, as summarized in Table V.
Within the stated errors, the measurements in the different momentum intervals are
consistent for both sets of calculations.

We adopt the results based on the more recent calculations (BLNP)~\cite{neubert_extract},
since they represent a more complete theoretical analysis of the full electron spectrum
and relate the SF to the HQE parameters extracted from inclusive $B \ra X_s \gamma$
and $B \ra X_c \ell \nu$ decays. We choose the SF parameters obtained from the combined
fit to moments of inclusive distributions measured by the \babar\ Collaboration rather
than the single most precise measurement of the SF parameters obtained from the recent
\babar\ measurement~\cite{babar_photons} of the semi-inclusive photon spectrum in
$B \ra X_s \gamma$ decays.
Assuming it is valid to combine the experimental and the estimated theoretical errors
in quadrature, and taking into account the fraction of the signal contained in this
interval, we conclude that the best measurement can be extracted from the largest
momentum interval, $2.0\,$ to $\,2.6 \gevc$. For this momentum interval the partial
branching fraction is
\begin{eqnarray}
& & \Delta {\cal B}(B \rightarrow X_u e \nu)= \\
& &   (0.572 \pm 0.041_{stat} \pm 0.065_{syst})\times 10^{-3}. \nonumber
\end{eqnarray}
Here the first error is statistical and the second is the total systematic error,
as listed in Table~\ref{table:t2}. In addition to the systematic uncertainty due to the
signal extraction, the normalization, and various small corrections, this error also
includes the observed dependence of the extracted signal on the choice of the SF
parameters. Based on the BLNP method, we obtain a total branching fraction of
\begin{eqnarray}
& & {\cal B}(B \rightarrow X_u e \nu) = \\
& &   (2.27 \pm 0.26_{exp} \,\,\,^{+0.33}_{-0.28}\,\,\,_{SF}
\pm 0.17_{theory})\times 10^{-3}, \nonumber
\end{eqnarray}
and
\begin{eqnarray}
& & |V_{ub}|= \\ 
& & (4.44 \pm 0.25_{exp} \,\,\,^{+0.42}_{-0.38}\,\,\,_{SF}
\pm 0.22_{theory}) \times 10^{-3}. \nonumber
\end{eqnarray}
Here the first error represents the total experimental
uncertainty, the second refers to the
uncertainty in the SF parameters from the combined fit to moments,
and the third
combines the stated theoretical uncertainties
in the extraction of $|V_{ub}|$, including
uncertainties from the subleading SFs, weak annihilation
effects, and various scale-matching uncertainties.
No additional uncertainty due to the theoretical assumption of quark-hadron
duality has been assigned.

The improvement in precision compared to earlier analyses of the lepton spectrum near
the kinematic endpoint can be attributed to improvements in
experimental techniques, to higher statistics, and in particular, to improved background
estimates, as well as significant advances in the theoretical understanding
of the SFs and extraction of the SF parameters from
inclusive spectra and moments. While earlier measurements were restricted to
lepton energies close to the kinematic endpoint for $B\ra X_c \ell \nu$ decays
at 2.3 \gevc and covered only 10\% of the $B\ra X_u \ell \nu$ spectrum, these and
other more recent measurements have been extended to lower momenta, including about
25\% of the spectrum, and thus have resulted in a significant reduction in the
theoretical uncertainties on $|V_{ub}|$.

The determination of $|V_{ub}|$ is currently limited primarily by our knowledge
of SF parameters. An approximate linear dependence of $|V_{ub}|$ on these parameters is
\begin{equation}
\frac{\Delta |V_{ub}|}{|V_{ub}|} = 
1.31 \frac{\Delta \bar{\Lambda}}{\bar{\Lambda}}
- 0.04 \frac{\Delta \mu_{\pi}^2}{\mu_{\pi}^2}.
\end{equation}
\noindent
for $\bar{\Lambda}=0.69 \gevcc$ and $\mu_{\pi}^2=0.21 \gev^2$. Thus the uncertainty
on the $b$-quark mass dominates. It should be noted that this dependence on
$\bar{\Lambda}$ is a factor of two smaller for measurements based on the DN calculations.

These results are in excellent agreement with earlier measurements of the
inclusive lepton spectrum at the \FourS\ resonance, but their overall precision
surpasses them~\cite{argus,cleo,cleo2,belle-vub}. The earlier results were based
on the DN calculations. We observe that for the same experimental input, i.e.\ the
same measured lepton and photon spectra, the extracted values of $|V_{ub}|$ based
on DN calculations agree very well with those based on BNLP calculations for the
various momentum ranges under study, even though the corresponding partial branching
fractions may differ by one standard deviation.

The results presented here are also comparable in precision to, and fully compatible
with, inclusive measurements recently published by the \babar~\cite{babar_elq2,babar_mxq2}
and Belle~\cite{belle_mxq2} Collaborations, based on two-dimensional distributions of
lepton energy, the momentum transfer squared and the hadronic mass, with SF parameters
extracted from $B\ra X_s \gamma$ and $B\ra X_c \ell \nu$ decays.

\section{ACKNOWLEDGMENTS}
\label{sec:acknowledgments}
We would like to thank the CLEO and Belle Collaborations for
providing detailed information on the extraction of 
the shape function parameters from the photon spectrum
in $b\to s \gamma$ transitions.
We are also indebted to M.\ Neubert and his co-authors
B.\ Lange, G.\ Paz, and S.\ Bosch
for providing us with detailed information on their calculations.
We are grateful for the 
extraordinary contributions of our \pep2\ colleagues in
achieving the excellent luminosity and machine conditions
that have made this work possible.
The success of this project also relies critically on the
expertise and dedication of the computing organizations that
support \babar.
The collaborating institutions wish to thank
SLAC for its support and the kind hospitality extended to them.
This work is supported by the
US Department of Energy
and National Science Foundation, the
Natural Sciences and Engineering Research Council (Canada),
Institute of High Energy Physics (China), the
Commissariat \`a l'Energie Atomique and
Institut National de Physique Nucl\'eaire et de Physique des Particules
(France), the
Bundesministerium f\"ur Bildung und Forschung and
Deutsche Forschungsgemeinschaft
(Germany), the
Istituto Nazionale di Fisica Nucleare (Italy),
the Foundation for Fundamental Research on Matter (The Netherlands),
the Research Council of Norway, the
Ministry of Science and Technology of the Russian Federation, and the
Particle Physics and Astronomy Research Council (United Kingdom).
Individuals have received support from
CONACyT (Mexico),
the A.\ P.\ Sloan Foundation,
the Research Corporation,
and the Alexander von Humboldt Foundation.


\begin{thebibliography}{99}
\bibitem{beta}
\babar\ Collaboration, B.\ Aubert {\em et al.},
Phys.\ Rev.\ Lett.\ {\bf 87}, 091801 (2001);
Phys.\ Rev.\ D {\bf 66}, 032003 (2002);
Phys.\ Rev.\ Lett.\ {\bf 89}, 201802 (2002).

\bibitem{phi1}
Belle Collaboration, K.\ Abe {\em et al.},
Phys.\ Rev.\ Lett.\ {\bf 87}, 091802 (2001);
Phys.\ Rev.\ D {\bf 66}, 032007 (2002);
Phys.\ Rev.\ D {\bf 66}, 071102 (2002).

\bibitem{sm}
M.\ Kobayashi and T.\ Maskawa, Prog.\ Theor.\ Phys.\ {\bf 49}, 652 (1973).

\bibitem{ope}
M.\ Shifman and M.\ Voloshin, Sov.\ J.\ Nucl.\ Phys.\ {\bf 41}, 120 (1985);
J.\ Chay, H.\ Georgi, and B.\ Grinstein, Phys.\ Lett.\ B {\bf 247}, 399 (1990);
I.\ I.\ Bigi and N.\ Uraltsev, Phys.\ Lett.\ B {\bf 280}, 271 (1992);
A.\ V.\ Manohar and M.\ B.\ Wise, Phys.\ Rev.\ D {\bf 49}, 1310 (1994);
B.\ Blok, L.\ Koyrakh, M.\ Shifman and A.\ I.\ Vainshtein,
Phys.\ Rev.\ D {\bf 49}, 3356 (1994).

\bibitem{neubert94}
M.\ Neubert, Phys.\ Rev.\ D {\bf 49}, 3392 (1994);
C.\ W.\ Bauer and A.\ V.\ Manohar, Phys.\ Rev.\ D {\bf 70}, 034024 (2004).


\bibitem{kolya94}
I.\ I.\ Bigi, M.\ A.\ Shifman, N.\ G.\ Uraltsev, and A.\ I.\ Vainshtein,
Int.\ J.\ Mod.\ Phys.\ A {\bf 9}, 2467 (1994).


\bibitem{neubert94a}
M.\ Neubert, Phys.\ Rev.\ D {\bf 49}, 4623 (1994).

\bibitem{argus}
ARGUS Collaboration, H.\ Albrecht {\em et al}.,
Phys.\ Lett.\ B {\bf 234}, 409 (1990);
Phys.\ Lett.\ B {\bf 255}, 297 (1991).

\bibitem{cleo}
CLEO Collaboration, R.\ Fulton {\em et al}., Phys.\ Rev.\ Lett.\ {\bf 64}, 16 (1990);
J.\ Bartelt {\em et al}., Phys.\ Rev.\ Lett.\ {\bf 71}, 4111 (1993).

\bibitem{cleo2}
CLEO Collaboration,
A.\ Bornheim {\em et al}., Phys.\ Rev.\ Lett.\ {\bf 88}, 231803 (2002).

\bibitem{ichep04}
\babar\ Collaboration, B.\ Aubert {\em et al.},
{\it Measurement of the Inclusive Electron Spectrum in Charmless Semileptonic B Decays,}
Contributions to ICHEP02, Amsterdam (2002), hep-ex/0207081;
\babar\ Collaboration, B.\ Aubert {\em et al.},
{\it Determination of the Partial Branching Fraction for $B \ra X_u \ell \nu$ and of $|V_{ub}|$ from
the Inclusive Electron Spectrum near the Kinematic Endpoint,}
Contribution to ICHEP04, Beijing (2004), hep-ex/0408075.

\bibitem{belle-vub}
Belle Collaboration,
A.\ Limosani {\em et al}., Phys.\ Lett.\ B {\bf 621}, 28 (2005).

\bibitem{l3-vub}
L3 Collaboration,
M.\ Acciarri {\em et al}., Phys.\ Lett.\ B {\bf 436}, 174 (1998).

\bibitem{aleph-vub}
ALEPH Collaboration,
R.\ Barate {\em et al}., Eur.\ Phys.\ J.\ C {\bf 6}, 555 (1999).

\bibitem{delphi-vub}
DELPHI Collaboration,
P.\ Abreu {\em et al}., Phys.\ Lett.\ B {\bf 478}, 14 (2000).

\bibitem{opal-vub}
OPAL Collaboration,
R.\ Barate {\em et al}., Eur.\ Phys.\ J.\ C {\bf 21}, 399 (2001).

\bibitem{dFN}
F.\ DeFazio, and M.\ Neubert, JHEP {\bf 9906}, 017 (1999).

\bibitem{kagan_neubert}
A.\ L.\ Kagan and M.\ Neubert, Eur.\ Phys.\ J.\ C {\bf 7}, 5 (1999).

\bibitem{blnp1}
S.\ W.\ Bosch, B.\ O.\ Lange, M.\ Neubert, and G.\ Paz,
Nucl.\ Phys.\ B {\bf 699}, 335 (2004).

\bibitem{blnp2}
M.\ Neubert,
Eur.\ Phys.\ J.\ {\bf C40}, 165 (2005).

\bibitem{blnp3}
S.\ W.\ Bosch, M.\ Neubert and G.\ Paz,
JHEP 0411, 073 (2004) 
\bibitem{blnp4}
M.\ Neubert,
{\it Impact of Four-Quark Shape Functions on Inclusive B Decay Spectra,}
hep-ph/0411027 (2004).

\bibitem{neubert_loops}
M.\ Neubert, Phys.\ Lett.\ {\bf B612}, 13 (2005) 
and private communication.

\bibitem{neubert_extract}
B.\ Lange, M.\ Neubert, and G.\ Paz,
{\it Theory of Charmless Inclusive B Decays and the Extraction of $|V_{ub}|$.}
hep-ph/0504071 (2005).

\bibitem{babar_photons}
\babar\ Collaboration, B.\ Aubert {\em et al.},
{\it Measurement of the $B \ra X_s \gamma$ Branching Fraction and Photon Spectrum from a Sum of
Exclusive Final States,}, submitted to PRD,
hep-ex/0508004 (2005).

\bibitem{babarvcb}
\babar\ Collaboration, B.\ Aubert {\em et al.},
Phys.\ Rev.\ Lett.\ {\bf 93}, 011803 (2004).

\bibitem{babarcf}
O.\ Buchm\"{u}ller, H.\ Fl\"{a}cher,
{\it Fits to Moment Measurements from $B \ra X_c \ell \nu$ and $B\ra X_s \gamma$
Decays Using Heavy Quark Expansions in the Kinetic Scheme,}
hep-ph/0507253 (2005).

\bibitem{detector}
\babar\ Collaboration, B.\ Aubert {\em et al.},
Nucl.\ Instr.\ Methods Phys.\ Res., Sect.\ A {\bf 479}, 1 (2002).

\bibitem{thorsten}
\babar\ Collaboration, B.\ Aubert {\em et al.},
Phys.\ Rev.\ D {\bf 67} 031101 (2003).

\bibitem{geant4}
GEANT4 Collaboration,
S.\ Agostinelli {\em et al.}, Nucl.\ Instr.\ Methods Phys.\ Res., Sect.\ A
{\bf 506}, 250 (2003).

\bibitem{photos}
E.\ Richter-Was, Phys.\ Lett.\ B {\bf 303}, 163 (1993).

\bibitem{pdg2002}
Particle Data Group,
K.\ Hagiwara , Phys.\ Rev.\ D {\bf 66}, 010001 (2002).

\bibitem{jetset}
T.\ Sj{\"o}strand, Comput.\ Phys.\ Commun.\ {\bf 82}, 74 (1994).

\bibitem{isgw2}
N.\ Isgur, D.\ Scora, B.\ Grinstein, and M.\ B.\ Wise, Phys.\ Rev.\ 
 D {\bf 39}, 799 (1989); 
D.\ Scora, N.~Isgur, Phys.\ Rev.\ D {\bf 52}, 2783 (1995).


\bibitem{hqet}
I.\ I.\ Bigi, M.\ Shifman, and N.\ G.\ Uraltsev, Annu.\ Rev.\ Nucl.\ Part.\ Sci.\ 
{\bf 47}, 591 (1997).

\bibitem{CLN}
I.\ Caprini, L.\ Lellouch, M.\ Neubert, Nucl.\ Phys.\ B {\bf 530}, 153 (1998).

\bibitem{GL}
B.\ Grinstein and Z.\ Ligeti, Phys.\ Lett.\ B {\bf 526}, 345 (2002).


\bibitem{bd_form}
CLEO Collaboration, J.\ Bartelt {\em et al.}, Phys.\ Rev.\ Lett.\ {\bf 82}, 3746 (1999).

\bibitem{bd_form2}
Belle Collaboration, K.\ Abe {\em et al.}, Phys.\ Lett.\ B {\bf 526}, 258 (2002).


\bibitem{babarff}
\babar\ Collaboration, B.\ Aubert {\em et al.},
{\it Measurement of the $B \ra D^*$ Form Factors in the Semileptonic Decay
$\bar{B} \ra D^{*+} e^- \bar{\nu}$ },
Contribution to ICHEP04, Beijing (2004), hep-ex/0409047.

\bibitem{gr}
J.\ L.\ Goity and W.\ Roberts, Phys.\ Rev.\ D {\bf 51}, 3459 (1995).

\bibitem{foxw}
G.\ C.\ Fox and S.\ Wolfram, Phys.\ Rev.\ Lett.\ {\bf 41}, 1581 (1978).

\bibitem{pdg2004}
Particle Data Group, 
S.\ Eidelman et al., Phys.\ Lett.\ B {\bf 592}, 1 (2004).

\bibitem{vub1}
N.\ Uraltsev, Int.\ J.\ Mod.\ Phys.\ A {\bf 14}, 4641 (1999),
and private communication (2004).

\bibitem{vub2}
A.\ H.\ Hoang, Z.\ Ligeti, and A.\ V.\ Manohar, Phys.\ Rev.\ D {\bf 59}, 074017 (1999).

\bibitem{vub3}
T.\ van Ritbergen, Phys.\ Lett.\ B {\bf 454}, 353 (1999).

\bibitem{private}
M.\ Neubert, private communication (2004).

\bibitem{cleo_photons}
CLEO Collaboration, S.\ Chen {\em et al.}, Phys.\ Rev.\ Lett.\ {\bf 87}, 251807 (2001).

\bibitem{belle_photons}
Belle Collaboration, P.\ Koppenburg {\em et al.},
Phys.\ Rev.\ Lett.\ {\bf 93}, 061803 (2004).

\bibitem{kinetic}
D.\ Benson, I.\ I.\ Bigi, and N.\ Uraltsev, Nucl.\ Phys.\ {\bf 710}, 371 (2005).

\bibitem{babar_elq2}
\babar\ Collaboration, B.\ Aubert {\em et al.}, Phys.\ Rev.\ Lett.\ {\bf 95} 111801 (2005).

\bibitem{babar_mxq2}
\babar\ Collaboration, B.\ Aubert {\em et al.},
{\it Measurement of the Partial Branching Fraction for Inclusive Charmless Semileptonic $B$ Decays and the Extraction of $|V_{ub}|$}, hep-ex/0507017,
Contribution to the Int.\ Symposium of Lepton-Photon Interactions, Uppsala (2005).

\bibitem{belle_mxq2}
Belle Collaboration, H.\ Kakuno {\em et al.},
Phys.\ Rev.\ Lett.\ {\bf 92}, 101801 (2004).

\end{thebibliography}
\end{document}